\documentclass[aps,prd]{revtex4}

\usepackage{amsmath}
\usepackage{amssymb}
\usepackage{bm}
\usepackage{graphicx}
\usepackage{multirow}

\usepackage{textcomp}

\allowdisplaybreaks[1]

\begin{document}

\title{Dispersive constraints on fermion masses}

\author{Hsiang-nan Li}
\affiliation{Institute of Physics, Academia Sinica,
Taipei, Taiwan 115, Republic of China}

\date{\today}

\begin{abstract}
We demonstrate that fermion masses in the Standard Model (SM) can be constrained by the dispersion 
relations obeyed by hadronic and semileptonic decay widths of a fictitious heavy quark $Q$ with an 
arbitrary mass. These relations, imposing stringent connections between high-mass 
and low-mass behaviors of decay widths, correlate a heavy quark mass and the chiral 
symmetry breaking scale. Given the known input from leading-order heavy quark expansion and a 
hadronic threshold for decay products, we solve for a physical heavy quark decay width. It is shown 
that the charm (bottom) quark mass $m_c= 1.35$ GeV ($m_b= 4.0$ GeV) can be determined by 
the dispersion relation for the $Q\to du\bar d$ ($Q\to c\bar ud$) decay with the threshold $2m_\pi$ 
($m_\pi+m_D$), where $m_\pi$ ($m_D$) denotes the pion ($D$ meson) mass. Requiring that the dispersion 
relation for the $Q\to su\bar d$ ($Q\to d\mu^+\nu_\mu$, $Q\to u\tau^-\bar\nu_\tau$) decay with the 
threshold $m_\pi+m_K$ ($m_\pi+m_\mu$, $m_\pi+m_\tau$) yields the same heavy quark mass, $m_K$ being 
the kaon mass, we obtain the strange quark (muon, $\tau$ lepton) mass $m_s= 0.12$ 
GeV ($m_\mu=0.11$ GeV, $m_\tau= 2.0$ GeV). Moreover, all the predicted decay widths 
corresponding to the above masses agree with the data. It is pointed out that our formalism is similar 
to QCD sum rules for probing resonance properties, and that the Pauli interference (weak 
annihilation) provides the higher-power effect necessary for establishing the solutions of the 
hadronic (semileptonic) decay widths. This work suggests that the parameters in the SM may not be  
free, but arranged properly to achieve internal dynamical consistency.


\end{abstract}


%
%
%

\maketitle

%
%
%

\section{INTRODUCTION}

How to understand the flavor structures of the Standard Model (SM), such as the hierarchy of quark 
and lepton masses, and the dramatically different quark and lepton mixing patterns, has been a 
long-term pursuit in particle physics. Many proposals have been made in the literature, which usually
rely on additional symmetries or interactions. For example, a flavor U(1) symmetry with a scalar 
field called flavon was introduced and then spontaneously broken in the Froggatt-Nielsen model 
\cite{FN79}. The aforementioned flavor structures were generated in various models with modular 
flavor symmetry \cite{Feruglio:2017spp}, which has attracted great attention
\cite{Du:2022lij,Petcov:2022fjf,Kikuchi:2023cap,Bree:2023ojl,Abe:2023ilq}. 
The smooth confinement mechanism without chiral symmetry breaking 
\cite{Seiberg:1994bz,Seiberg:1994pq,Razamat:2020kyf} was implemented to explain the small Yukawa 
couplings of the first- and second-family fermions, which are composites of ultraviolet 
fields, while the third-family fermions are elementary to be consistent with the $O(1)$ top Yukawa 
coupling \cite{Hamada:2022ino}. Alternative scenarios, which do not resort to 
symmetries but to localizations of fermions in extra dimensions \cite{Arkani-Hamed:1999ylh} or to the 
clockwork mechanism \cite{Giudice:2016yja,Craig:2017cda}, were also attempted. We will demonstrate,
without any new ingredients beyond the SM, that at least the fermion masses around the GeV scale, 
including the strange quark, charm quark, bottom quark, muon and $\tau$ lepton masses, may be 
understood through the internal consistency of SM dynamics.

We performed a dispersive analysis on the $D$ meson mixing recently \cite{Li:2022jxc}, starting 
with the transition matrix elements, which contain the flavor-changing four-quark operators in 
effective weak Hamiltonians, for a fictitious $D$ meson of an arbitrary mass. Surprisingly, the 
solution to the dispersion relation obeyed by the transition matrix elements appears at the physical 
$D$ meson mass $m_D$. The emergence of the scale $m_D$ in the dispersive analysis inspires 
two speculations. First, an appropriate correlation function defined by the four-quark effective 
operators can be employed to establish the mass of a decaying heavy meson, similar to the 
determination of a light resonance mass that has been achieved extensively in QCD sum rules 
\cite{SVZ}. The heavy quark invariant mass squared in the dispersive analysis plays the role of the 
invariant momentum squared injected into a current operator in sum rules. The heavy quark expansion 
(HQE) for the evaluation of a heavy meson matrix element corresponds to the operator product 
expansion in sum rules. The distinction arises from the operators (four-quark operators 
in the former and currents in the latter) and the external states (heavy meson states in the former 
and the vacuum state in the latter), which sandwich the operators, for defining a correlation 
function. Second, the absorptive piece of the transition matrix elements receives contributions from
various fictitious $D$ meson decay channels. It is then likely to correlate a heavy quark mass and 
masses of its light decay products, which originate from the chiral symmetry breaking in QCD, through 
the dispersion relation. The above two speculations suggest the possibility of addressing at least 
partial flavor structure of the SM in the dispersive analysis on inclusive heavy quark decays.


We will study a hadronic matrix element of the four-quark effective operators, whose absorptive piece 
defines a decay width of a fictitious heavy quark $Q$ with an arbitrary mass $m_Q$. The associated 
dispersion relation is derived, which imposes a stringent connection between the high-mass and 
low-mass behaviors of the decay width, and then solved directly, namely, treated as an inverse 
problem \cite{Li:2020xrz}. It has been proved rigorously \cite{Xiong:2022uwj} that a unique solution 
exists for this type of integral equation, when boundary conditions are specified. The decay width 
at large $m_Q$ approaches its HQE in powers of $1/m_Q$, which has been known 
to accommodate the observed $B$ and $B_s$ meson lifetimes well. It ought to vanish at a hadronic 
threshold, which originates from the chiral symmetry breaking for light quarks. This threshold depends 
on final states, taking, for instance, $2m_\pi$ with the pion mass $m_\pi$ for the channel involving 
only up and down quarks. Following the proposal in \cite{Li:2021gsx}, we scale the dispersion relation 
by changing $m_Q^2=u\Lambda$ into a dimensionless variable $u$, which introduces the arbitrary
scale $\Lambda$. A solution to the dispersion relation must not be affected by this artificial 
variable change. It turns out, given the boundary conditions at infinity and the threshold, that 
only when $m_Q$ takes a specific value, can the stability with respect to the variation of $\Lambda$  
be realized. 


It will be shown that the above specific $m_Q$ solved from the dispersion relation does coincide with 
the mass of a physical heavy (charm or bottom) quark. We first perform the dispersive analysis of the 
hadronic decay $Q\to du\bar d$, regarding the final-state up and down quarks as being massless, with 
the leading-order (LO) HQE input and the threshold $2m_\pi$. The pion mass $m_\pi$ is a result of the 
chiral symmetry breaking in QCD as stated before. The charm quark mass $m_c=1.35$ GeV is then inferred, 
close to the value extracted from measured $D$ meson lifetimes \cite{Gratrex:2022xpm}. 
We then proceed to the investigation of the 
$Q\to su\bar d$ decay with the threshold $m_\pi+m_K$, $m_K$ being the kaon mass. It is encouraging 
to find that the strange quark mass $m_s$ must take a value around $m_s=0.12$ GeV in order to produce 
the same charm quark mass from the $Q\to du\bar d$ analysis. The relation between $m_s$ and 
$m_K$ is governed by strong interaction as verified in lattice QCD \cite{Blum:1999xi} and sum rules 
\cite{Dominguez:2007my}. The above observations indicate the internal consistency among the 
scales $m_s$, $m_\pi$, $m_K$ and $m_c$, which characterize strong and weak dynamics in the
SM. Moreover, the solved $c\to s\bar ud$ decay width $3.3\times 10^{-13}$ GeV is reasonable,
compared with the data of the Cabibbo-favored inclusive $D^+$ meson decay modes \cite{HFLAV:2022pwe}. 
The semileptonic decay $Q\to d\mu^+\nu_\mu$ is examined in a similar manner by considering the 
threshold $m_\pi+m_\mu$, $m_\mu$ being the muon mass. It is noticed that the dispersive constraint on 
the muon mass is quite rigid, which must take the value $m_\mu=0.11$ GeV in order to generate the same 
charm quark mass. 

We repeat the dispersive analysis on the width of the hadronic decay $Q\to c\bar ud$ by 
inputting the charm quark mass $m_c=1.35$ GeV derived previously and the threshold $m_\pi+m_D$, which 
leads to the bottom quark mass $m_b= 4.0$ GeV, consistent with the value extracted from measured $B$ 
meson lifetimes \cite{Cheng:2018rkz}. The relation between $m_c$ and $m_D$ can be 
deduced in QCD, so no a priori information on the bottom 
quark is introduced. The predicted decay width at $m_b=4.0$ GeV also agrees with the $b\to c\bar ud$ 
inclusive data in \cite{HFLAV:2022pwe}. Demanding that the dispersion relation for the semileptonic 
decay $Q\to u\tau^-\bar\nu_\tau$ with the threshold $m_\pi+m_\tau$ gives the same bottom quark 
mass, we fix the $\tau$ lepton mass $m_\tau= 2.0$ GeV. To sum up, the above fermion masses can be 
determined, starting from massless up and down quarks, by the dispersion relations which correlate 
ultraviolet and infrared behaviors of meson weak decays. We point out that the solution for a decay 
width reduces to the HQE input, once the chiral symmetry is restored, and no constraint on fermion 
masses can be imposed. It is stressed that the Pauli interference (weak annihilation) provides the 
higher-power effect essential for establishing a solution of the hadronic (semileptonic) decay width. 
For this reason, the semileptonic decay into the $e\nu_e$ final state does not serve the purpose of 
constraining the involved fermion masses efficiently because of the helicity suppression on weak 
annihilation.


The rest of the paper is organized as follows. We construct the dispersion relation obeyed by the 
absorptive piece of a heavy meson matrix element of the four-quark effective operators in Sec.~II. 
The $Q\to du\bar d$ case with the LO HQE input for massless decay products is explored in detail to 
illustrate our formalism. The equation for specifying the physical heavy quark mass is presented as a 
consequence of the scale invariance in the arbitrary $\Lambda$, which generates the charm quark mass 
$m_c$. We study the $Q\to su\bar d$, $Q\to c\bar ud$, $Q\to d\mu^+\nu_\mu$ and 
$Q\to u\tau^-\bar\nu_\tau$ decays into massive final states for fixing the masses $m_s$, $m_b$, 
$m_\mu$ and $m_\tau$, respectively, in Sec.~III. It is corroborated that the solved decay widths 
exhibit apparent stability under the variation of $\Lambda$, and match the data satisfactorily. 
Section~IV contains the conclusion and outlook.


\section{DISPERSIVE CONSTRAINTS}

The nonperturbative approach based on dispersion relations for physical observables
was proposed in \cite{Li:2020xrz}, and then applied to the constraint on the hadronic vacuum 
polarization contribution to the muon anomalous magnetic moment \cite{Li:2020fiz}, to the 
reformulation of QCD sum rules for extracting properties of the series of $\rho$ resonances 
\cite{Li:2020ejs}, glueball masses \cite{Li:2021gsx} and the pion light-cone distribution 
amplitude \cite{Li:2022qul}, and to the explanation of the large observed $D$ meson mixing parameters
\cite{Li:2022jxc}. Here we will extend it to the analysis of heavy quark decay widths and 
demonstrate that the involved fermion masses can be constrained, as the hadronic thresholds are 
introduced into the relevant dispersion relations. We concentrate on the $Q\to du\bar d$ case 
with massless final-state quarks and the determination of the charm quark mass in this section.

\subsection{Dispersion Relation}

Consider the analytical correlation function $\Pi(m_Q)$ for a heavy meson $H_Q$ of the mass 
$m_{H_Q}$ formed by the fictitious heavy quark $Q$, which is defined by the matrix element 
\begin{eqnarray}
\Pi(m_Q) \equiv \frac{1}{2m_{H_Q}(m_Q)}\langle H_Q|{\cal T}|H_Q\rangle
=M(m_Q)+i\Gamma(m_Q).
\end{eqnarray}
The transition operator ${\cal T}$ is written as
\begin{eqnarray}
{\cal T}=i\int d^4x T\left[{\cal H}^{\dagger}_{\rm eff}(x){\cal H}_{\rm eff}(0)\right],\label{eff}
\end{eqnarray}
where ${\cal H}_{\rm eff}$ denotes the $\Delta Q=1$ effective weak Hamiltonian. The functions 
$M(m_Q)$ and $\Gamma(m_Q)$ represent the dispersive and absorptive pieces, respectively. It has 
been observed that power corrections are crucial for establishing a resonance solution in QCD sum 
rules \cite{Li:2020ejs}. Hence, we stick to those heavy meson decays, which contain sizable 
higher-power corrections. It is known that the $D^0$ 
meson lifetime receives a dimension-six contribution only from the $W$-exchange topology, which 
suffers chiral suppression under the vacuum insertion approximation \cite{Lenz:2013aua}. 
It is possible to constrain the strange quark mass through the introduction of the $m_s$-dependent
threshold into the dispersion relation, so we will not discuss $D_s$ meson decays. We thus analyze 
$D^+$ meson decays below, whose widths in the HQE have been available in the literature.

The inclusive hadronic (semileptonic) decay width of a charged heavy meson is written, in the HQE, as 
\cite{KSUV,CGG,Cheng:2018rkz}
\begin{eqnarray}
\Gamma^{\rm HQE}_{h(s)}(m_Q) = \frac{G_F^2|V_{\rm CKM}|^2m_Q^5}{192\pi^3}
\left[c_{h(s)}^{(3)}\left(1-\frac{\mu_\pi^2-\mu_G^2}{2m_Q^2}\right)
+2c_{h(s)}^{(5)}\frac{\mu_G^2}{m_Q^2}
+\Gamma^{(6)+(7)}_{h(s)}(m_Q) \right],
\label{hq}
\end{eqnarray}
where $G_F$ is the Fermi constant and $V_{\rm CKM}$ represents the Cabibbo-Kobayashi-Maskawa (CKM) 
factor. The heavy-quark-effective-theory (HQET) parameters $\mu_\pi^2$ and $\mu_G^2$ are defined by 
the matrix elements $\langle H_Q|\bar QQ|H_Q\rangle$ and 
$\langle H_Q|\bar Q\sigma_{\mu\nu}G^{\mu\nu}Q|H_Q\rangle$,
respectively, with the gluon field strength $G^{\mu\nu}$. The significance of the Darwin term is still 
uncertain, which depends on how it is extracted \cite{Lenz:2022rbq}. Here we do not take it into 
account according to \cite{Bernlochner:2022ucr}: a small Darwin contribution, consistent with zero 
basically, yields a better fit to the measured lifetime ratio $\tau(B_s)/\tau(B_d)$. The coefficient 
functions of the penguin operators vanish at LO in the strong coupling $\alpha_s$, the precision we 
will work on. It has been postulated that the HQE can be truncated to a good approximation after the 
dimension-seven terms \cite{Lenz:2013aua}, as done in Eq.~(\ref{hq}).

For our purpose, it suffices to adopt the LO expressions for the hard coefficients \cite{Bigi:1992su} 
\begin{eqnarray}
& c_h^{(3)}=A_0(\mu)I_0(x_q),\;\;\;\;
&c_h^{(5)}= -A_0(\mu)I_1(x_q)-A_2(\mu)I_2(x_q),\nonumber\\
& c_s^{(3)}=I_0(x_\ell),\;\;\;\;&c_s^{(5)}= -I_1(x_\ell),
\end{eqnarray}
with the ratios $x_q=m_q^2/m_Q^2$ and $x_\ell=m_\ell^2/m_Q^2$, $m_q$ ($m_\ell$) being the mass of 
a final-state quark $q=s$ or $c$ (charged lepton $\ell=\mu$ or $\tau$), which will be touched 
on in the next section. The functions $A_i$ depend on the Wilson coefficients $C_{1,2}$,
\begin{eqnarray}
A_0=N_cC_1^2+ N_cC_2^2 +2C_1C_2,\;\;\;\;A_2=8C_1C_2.
\end{eqnarray}
The functions $I_i$ are given by \cite{Falk:1994gw,Mannel:2017jfk}
\begin{eqnarray}
I_0(x) &= &1-8x+8x^3-x^4-12x^2\ln x,\nonumber\\
I_1(x) &=& (1-x)^4,\;\;\;\;I_2(x)=(1-x)^3.
\end{eqnarray}
The next-to-leading-order (NLO) corrections to 
$c^{(3)}$ can be found in Refs.~\cite{QP83,N89,Bagan:1994zd,Bagan:1994qw,Bagan:1995yf,Falk:1994gw}. 



The term $\Gamma^{(6)+(7)}_{h(s)}(m_Q^2)$ combines the dimension-six and -seven hard spectator 
contributions, including that from the Pauli interference (weak annihilation). The coefficient 
functions of the dimension-six operators for the hadronic decays were evaluated to LO in 
Refs. \cite{Neubert:1996we,Uraltsev:1996ta,Beneke:1996xe}, and to NLO in 
Refs. \cite{Franco:2002fc,Beneke:2002rj}. The dimension-seven contributions were computed only
partially to LO in \cite{Lenz:2013aua,Cheng:2018rkz,King:2021xqp,Gratrex:2022xpm}.
The LO results are summarized as
\begin{eqnarray}
\Gamma_h^{(6)+(7)}(m_Q) &=& 16\pi^2
\frac{f_{H_Q}^2m_{H_Q}}{m_Q^3}(C_1^2+C_2^2+2N_cC_1C_2)(1-x_q)^2
\left[1-\left(\frac{1+x_q}{1-x_q}+\frac{1}{2}\right)\frac{2\bar\Lambda}{m_Q}
\right],\label{67}\\
\Gamma_s^{(6)+(7)}(m_Q)&=&16\pi^2\frac{f_{H_Q}^2m_{H_Q}}{m_Q^3}x_\ell(1-x_\ell)
\left[\frac{3}{2}(1-x_\ell)
-3x_\ell(1-2x_\ell)\frac{2\bar\Lambda}{m_Q}\right],\label{68}
\end{eqnarray}
with the heavy meson decay constant $f_{H_Q}$ and the binding energy $\bar\Lambda=m_{H_Q}-m_Q$. 
The vacuum insertion approximations for the involved 
hadronic matrix elements have been applied to simplify the expressions, and the weak annihilation 
contribution to the hadronic decay width is negligible compared with the Pauli interference one. 
Notice the two-body phase-space enhancement factor $16\pi^2$ relative to the three-body phase 
space. This clarifies why the spectator effects, despite being power suppressed, are important. 



\begin{figure}
\begin{center}
\includegraphics[scale=0.35]{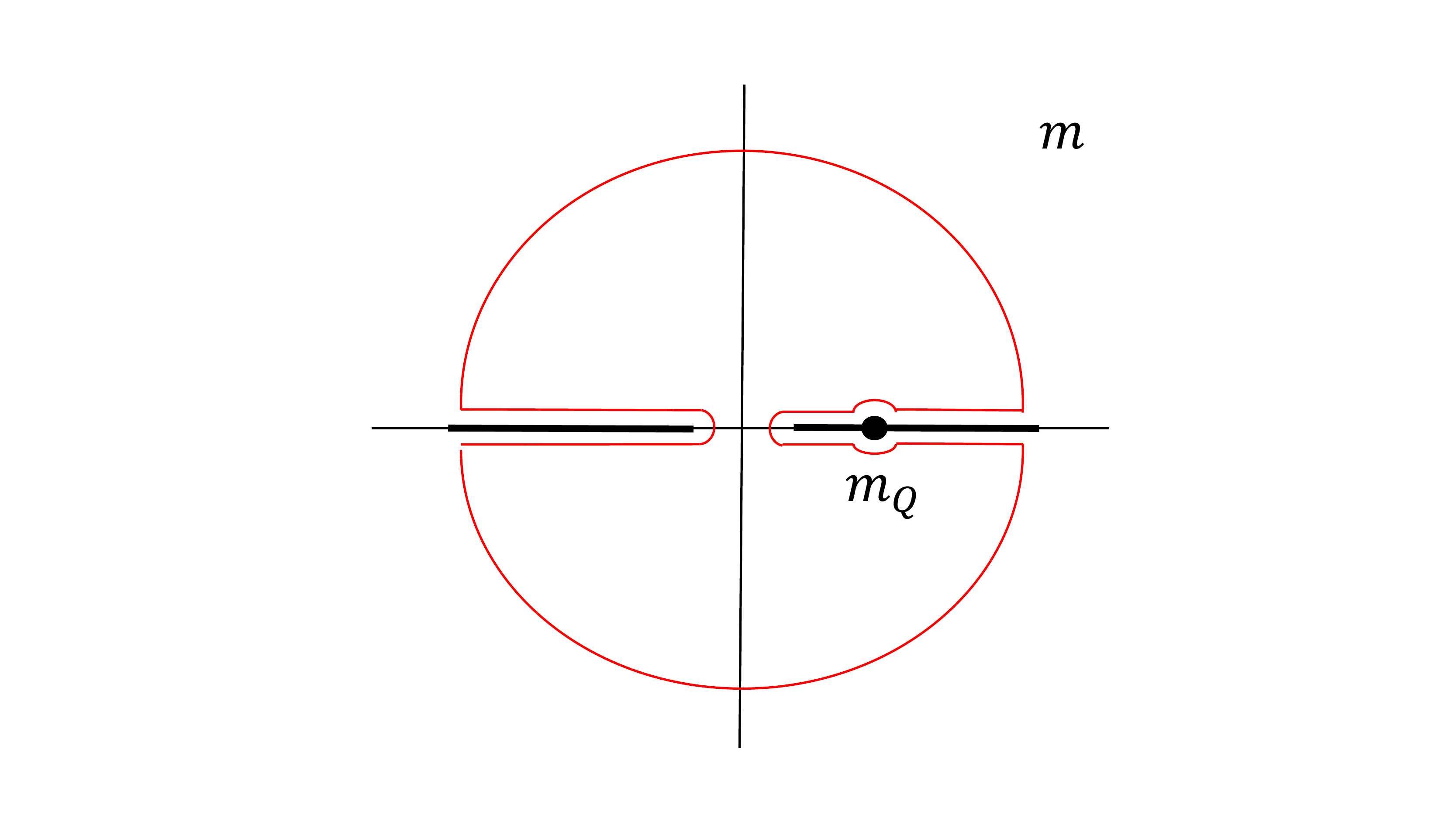}
\caption{\label{fig1}
Contour for Eq.~(\ref{con}), where the thick lines represent the branch cuts.}
\end{center}
\end{figure}

Since the HQE result contains terms in various powers of $1/m_Q$, instead of in $1/m_Q^2$, the 
construction of a dispersion relation begins with the contour integration in the complex $m_Q$ 
plane \cite{Falk:2004wg}, instead of the $m_Q^2$ plane \cite{Li:2022jxc}, which possesses different 
branching cuts. We have the identity
\begin{eqnarray}
\frac{1}{2\pi i}\oint dm\frac{\Pi(m)}{m_Q-m}=0,\label{con}
\end{eqnarray}
in which the contour consists of two pieces of horizontal lines above and below the branch cut
along the positive real axis, two pieces of horizontal lines above and below the branch cut
along the negative real axis, a small circle around the pole $m=m_Q$ located on the 
positive real axis and a circle $C_R$ of the large radius $R$ as depicted in Fig.~\ref{fig1}. 
The radius $R$ should not exceed the $W$ boson mass $m_W$ in order to validate the use of the  
effective weak Hamiltonians in Eq.~(\ref{eff}). The integral in Eq.~(\ref{con}) vanishes, for
the contour encloses only unphysical regions without poles. 

The contribution along the small clockwise circle yields $M(m_Q)$, and those from the four pieces 
of horizontal lines lead to the dispersive integrals of $\Gamma(m)$. Equation~(\ref{con}) 
becomes
\begin{eqnarray}
M(m_Q)=\frac{1}{\pi}\int_{m_F}^R dm\frac{\Gamma(m)}{m_Q-m}
-\frac{1}{\pi}\int^{-m_F}_{-R} dm\frac{\Gamma(m)}{m_Q-m}
+\frac{1}{2\pi i}\int_{C_R} dm\frac{\Pi^{\rm HQE}(m)}{m_Q-m},\label{ij}
\end{eqnarray}
where the hadronic threshold $m_F$ sums the masses in the lightest final state. The unknown function 
$\Gamma(m)$ acquires nonperturbative contributions from the small $m$ region, where the chiral 
symmetry is broken. It is the reason why the threshold $m_F$, which is dynamically generated and 
of order of the QCD scale $\Lambda_{\rm QCD}$, appears in Eq.~(\ref{ij}). The integrand $\Pi(m)$, 
taking values along the large counterclockwise circle $C_R$, can be reliably replaced by the 
perturbative one $\Pi^{\rm HQE}(m)$; as stated before, the HQE accounts for the measured
$B$ meson lifetimes well.


The dispersive part $M^{\rm HQE}(m_Q)$ and the absorptive part $\Gamma^{\rm HQE}(m_Q)$ from the
HQE respect the dispersion relation,
\begin{eqnarray}
M^{\rm HQE}(m_Q)=\frac{1}{\pi}\int_{0}^R dm\frac{\Gamma^{\rm HQE}(m)}{m_Q-m}
-\frac{1}{\pi}\int^{0}_{-R} dm\frac{\Gamma^{\rm HQE}(m)}{m_Q-m}
+\frac{1}{2\pi i}\int_{C_R} dm\frac{\Pi^{\rm HQE}(m)}{m_Q-m},\label{ope}
\end{eqnarray}
where the thresholds for the first two integrals on the right-hand side have been set to zero 
for the massless up and down quarks with $m_q=0$. There is no pole at $m_Q=0$ up to the power 
shown in Eq.~(\ref{hq}). Though $\Gamma^{\rm HQE}(m)$ does not describe the low-mass
behavior of a physical decay width correctly, Eq.~(\ref{ope}) holds simply owing to the 
analyticity of the perturbative quark-level calculation. We equate $M(m_Q)$ and $M^{\rm HQE}(m_Q)$, 
i.e., Eqs.~(\ref{ij}) and (\ref{ope}) at large enough $m_Q\gg m_F$, arriving at
\begin{eqnarray}
\int_{m_F}^R\frac{\Gamma(m)}{m_Q-m}dm-
\int^{-m_F}_{-R}\frac{\Gamma(m)}{m_Q-m}dm=
\int_{0}^R\frac{\Gamma^{\rm HQE}(m)}{m_Q-m}dm-
\int^{0}_{-R}\frac{\Gamma^{\rm HQE}(m)}{m_Q-m}dm,\label{ij1}
\end{eqnarray} 
where the contributions from the large circle $C_R$ on the two sides have been canceled.


\subsection{Solution of Decay Width}



We decompose the HQE hadronic width $\Gamma_h^{\rm HQE}(m)$ into the pieces, which are even and odd 
in powers of $m$, $\Gamma_h^{\rm HQE}(m)=\Gamma_e^{\rm HQE}(m)+\Gamma_o^{\rm HQE}(m)$, and 
the unknown function into $\Gamma_h(m)=\Gamma_e(m)+\Gamma_o(m)$ accordingly. For the even piece, the 
variable change $m\to -m$ applied to the second integrals on both sides of Eq.~(\ref{ij1}) results in
\begin{eqnarray}
\int_{m_F^2}^{R^2}\frac{\Gamma_e(m)}{m_Q^2-m^{2}}dm^{2}=
\int_{0}^{R^2}\frac{\Gamma_e^{\rm HQE}(m)}{m_Q^2-m^{2}}dm^{2}.
\end{eqnarray}
Moving the integrand on the right-hand side to the left-hand side, and regarding it as a 
subtraction term, we get
\begin{eqnarray}
\int_{0}^\infty\frac{\Delta\Gamma_e(m)}{m_Q^2-m^2}dm^2=0.\label{ge}
\end{eqnarray}
The subtracted unknown function $\Delta\Gamma_e(m)\equiv\Gamma_e(m)-\Gamma_e^{\rm HQE}(m)$ is fixed 
to $-\Gamma_e^{\rm HQE}(m)$ in the interval $(0,m_F)$ of $m$, and approaches zero at large $m$,
because of $\Gamma_e(m)\to\Gamma_e^{\rm HQE}(m)$ in this limit. Though $R$ should not exceed $m_W$,
we extend it to infinity owing to the diminishing of $\Delta\Gamma_e(m)$ at large $m$.


For the odd piece in $m$, the variable change $m\to -m$ applied to 
the second integrals on both sides of Eq.~(\ref{ij1}) gives
\begin{eqnarray}
m_Q\int_{m_F^2}^{R^2}\frac{\Gamma_o(m)}{m(m_Q^2-m^{2})}dm^{2}=
m_Q\int_{0}^{R^2}\frac{\Gamma_o^{\rm HQE}(m)}{m(m_Q^2-m^{2})}dm^{2}.
\end{eqnarray}
Moving the integrand on the right-hand side to the left-hand side leads to
\begin{eqnarray}
\int_{0}^\infty\frac{\Delta\Gamma_o(m)}{m(m_Q^2-m^{2})}dm^{2}=0.\label{go}
\end{eqnarray}
The subtracted unknown function $\Delta\Gamma_o(m)\equiv\Gamma_o(m)-\Gamma_o^{\rm HQE}(m)$ 
is fixed to $-\Gamma_o^{\rm HQE}(m)$ in the interval $(0,m_F)$, and approaches zero at 
large $m$. The implication of Eqs.~(\ref{ge}) and (\ref{go}) will be probed below. Since they 
hold for an arbitrary large scale $m_Q$, they impose a stringent correlation between the 
nonperturbative behavior of $\Gamma_h(m)$ at low mass and the perturbative behavior of 
$\Gamma_h(m)\approx \Gamma_h^{\rm HQE}(m)$ at high mass, thus among the relevant mass scales. It is 
obvious that the trivial solutions $\Delta\Gamma_{e,o}(m)=0$, i.e., $\Gamma_h(m)=\Gamma_h^{\rm HQE}(m)$ 
exist as $m_F\to 0$. In other words, there will be no constraint on fermion masses in the absence 
of the chiral symmetry breaking. We emphasize that Eqs.~(\ref{ge}) and (\ref{go}) hold for each decay 
channel of $H_Q$, because the associated CKM factor can vary independently from a mathematical 
point of view.


We change $m_Q^2$ and $m^2$ in Eq.~(\ref{ge}) into the dimensionless variables
$u$ and $v$ via $m_Q^2=u\Lambda$ and $m^2=v\Lambda$, respectively, obtaining
\begin{eqnarray}
\int_{0}^\infty dv\frac{\Delta\Gamma_e(v)}{u-v}=0.\label{i2}
\end{eqnarray}
The purpose of introducing the arbitrary scale $\Lambda$ will become clear shortly. Viewing the fact 
that $\Delta\Gamma_e(v)$ decreases at large $v$, and the major contribution to Eq.~(\ref{i2}) arises 
from the region with finite $v$, we are allowed to expand Eq.~(\ref{i2}) into a power series in $1/u$ 
for sufficiently large $u$ by inserting
\begin{eqnarray}
\frac{1}{u-v}=\sum_{i=1}^\infty \frac{v^{i-1}}{u^i}.
\label{ep}
\end{eqnarray}
Equation~(\ref{i2}) then demands a vanishing coefficient for every power of $1/u$. 

We start with the case of $N$ vanishing coefficients,
\begin{eqnarray}
\int_{0}^\infty dvv^{i-1}\Delta\Gamma_e(v)=0,\;\;\;\;i=1,2,3\cdots,N,\label{i3}
\end{eqnarray}
where $N$ is a large integer, such that Eq.~(\ref{i2}) is valid up to 
negligible corrections down by a power $1/u^{N+1}$. It hints that $\Delta\Gamma_e(v)$ can be expanded 
in terms of the generalized Laguerre polynomials $L_j^{(\alpha)}(v)$ with the support $[0,\infty)$, 
which respect the orthogonality condition
\begin{eqnarray}
\int_0^\infty v^\alpha e^{-v}L_i^{(\alpha)}(v)L_j^{(\alpha)}(v)dv
=\frac{\Gamma(i+\alpha+1)}{i!}\delta_{ij}.
 \label{ort}
\end{eqnarray}  
The first $N$ polynomials $L_{0}^{(\alpha)}(v)$, $L_{1}^{(\alpha)}(v)$, $L_{2}^{(\alpha)}(v)$, 
$\cdots$, $L_{N-1}^{(\alpha)}(v)$ are composed of the terms $1$, $v$, $v^2$, $\cdots$, $v^{N-1}$ 
appearing in Eq.~(\ref{i3}). Therefore, the expansion of $\Delta\Gamma_e(v)$ contains 
the polynomials with degrees $j$ not lower than $N$,
\begin{eqnarray}
\Delta \Gamma_e(v)=\sum_{j=N}^{N'} a_{j}v^\alpha e^{-v}L_{j}^{(\alpha)}(v),\;\;\;\;
N'>N,\label{d0}
\end{eqnarray}
with a set of unknown coefficients $a_{j}$. The highest degree $N'$ can be fixed by the initial
condition $\Delta \Gamma_e(v)=-\Gamma_e^{\rm HQE}(v)$ in the interval $(0,m_F^2/\Lambda)$ of $v$. 
Since $-\Gamma_e^{\rm HQE}(v)$ is a smooth function, $N'$ needs not be infinite.

A generalized Laguerre polynomial takes the approximate form for a large $j$ \cite{BBC}
\begin{eqnarray}
L_j^{(\alpha)}(v)\approx j^{\alpha/2}v^{-\alpha/2}e^{v/2}J_\alpha(2\sqrt{jv}),\label{Ln}
\end{eqnarray}
up to corrections of $1/\sqrt{j}$, $J_\alpha$ being a Bessel function of the first kind. 
Equation~(\ref{d0}) becomes
\begin{eqnarray}
\Delta \Gamma_e(m)
\approx\sum_{j=N}^{N'} a_{j}\sqrt{\frac{jm^2}{\Lambda}}^{\alpha} e^{-m^2/(2\Lambda)}
J_\alpha\left(2\sqrt{\frac{jm^2}{\Lambda}}\right),
\label{d1}
\end{eqnarray}
where the variable $v$ has been written as $m^2/\Lambda$ explicitly. Defining the scaling variable 
$\omega\equiv\sqrt{N/\Lambda}$, we have the approximation 
$N'/\Lambda=\omega^2+(N'-N)/N\approx \omega^2$ for a finite $N'-N$. Equation~(\ref{d1}) then reduces to
\begin{eqnarray}
\Delta \Gamma_e(m)\approx
y_e(\omega m)^{\alpha} J_\alpha\left(2\omega m\right),
\label{d2}
\end{eqnarray}
where the common Bessel functions $J_\alpha(2\sqrt{jm^2/\Lambda})\approx J_\alpha(2\omega m)$ for 
$j=N,N+1,\cdots,N'$ have been factored out, and the sum of the unknown coefficients 
$\sum_{j=N}^{N'} a_{j}$ has been denoted by $y_e$. The exponential suppression factor 
$e^{-s/(2\Lambda)}=e^{-\omega^2 m^2/(2N)}$ has been replaced by unity for a large $N$ in the 
region with finite $m$ and $\omega$, which we are interested in (see the next subsection). The 
correction to this replacement is of power $1/N$, smaller than that to Eq.~(\ref{Ln}). 

We replicate the above procedure for Eq.~(\ref{go}), deriving
\begin{eqnarray}
\frac{\Delta \Gamma_o(m)}{m}\approx
y_o(\omega m)^{\alpha} J_\alpha\left(2\omega m\right),\label{o2}
\end{eqnarray}
with the same index $\alpha$ as seen in the next subsection. A solution of the $Q\to du\bar d$ decay 
width is thus expressed, in terms of a single Bessel function, as
\begin{eqnarray}
\Delta \Gamma_h(m_Q)=\Delta \Gamma_e(m_Q)+\Delta \Gamma_o(m_Q)\approx
y_e\left(1+\frac{y_o}{y_e} m_Q\right)(\omega m_Q)^{\alpha} J_\alpha\left(2\omega m_Q\right).
\label{do2}
\end{eqnarray}
We stress that a solution to the dispersion relation should be insensitive to the variation 
of the arbitrary scale $\Lambda$, which is introduced via the artificial variable changes. 
The variation of $\Lambda$ is translated into that of $\omega$. To explain how to realize
this insensitivity, we make a Taylor expansion of $\Delta \Gamma_h(m_Q)$,
\begin{eqnarray}
\Delta \Gamma_h(m_Q)=\Delta \Gamma_h(m_Q)|_{\omega=\bar\omega}+
\frac{d\Delta \Gamma_h(m_Q)}{d\omega}\Big|_{\omega=\bar\omega}(\omega-\bar\omega)+
\frac{1}{2}\frac{d^2\Delta \Gamma_h(m_Q)}{d\omega^2}\Big|_{\omega=\bar\omega}
(\omega-\bar\omega)^2+\cdots,\label{ta}
\end{eqnarray}
where the constant parameter $\bar\omega$, together with $\alpha$, $y_e$ and $y_o$, 
are determined via the fit of the first term $\Delta \Gamma_h(m_Q)|_{\omega=\bar\omega}$ to 
$-\Gamma_h^{\rm HQE}(m_Q)$ in the interval $(0,m_F)$ of $m_Q$.

The insensitivity to the scaling variable $\omega$ requires the
vanishing of the first derivative in Eq.~(\ref{ta}),
\begin{eqnarray}
\frac{d\Delta \Gamma_h(m_Q)}{d\omega}\Big|_{\omega=\bar\omega}=0,\label{dd1}
\end{eqnarray}
from which roots of $m_Q$ are solved. Furthermore, the second derivative 
$d^2\Delta \Gamma_h(m_Q)/d\omega^2|_{\omega=\bar\omega}$ should be minimal to maximize the stability 
window around $\bar\omega$, in which $\Delta\Gamma_h(m_Q)$ is almost independent of the variation of 
$\omega$. Because the HQE result is independent of $\omega$, the stability of $\Delta\Gamma_h(m_Q)$ 
is equivalent to the stability of the decay width $\Gamma_h(m_Q)$. It will be shown that only when $m_Q$ 
takes a specific value can the above requirements be met. Equation~(\ref{do2}) with this specific 
$m_Q$ establishes a solution to the dispersion relation in Eq.~(\ref{ij1}) with the initial condition 
in the interval $(0,m_F)$ of $m_Q$. This specific $m_Q$ will be identified as the physical heavy quark 
mass, at which the corresponding $\Gamma_{h}=\Delta \Gamma_h+\Gamma_e^{\rm HQE}$ represents our 
prediction for the considered decay width. Once a solution for the decay width is found, the degree 
$N$ for the polynomial expansion in Eq.~(\ref{d0}) can be pushed, together with the scale 
$\Lambda$, to arbitrarily large values by keeping the scaling variable $\omega=\sqrt{N/\Lambda}$ within 
the stability window. Then all the arguments based on the large $N$ scenario, including the neglect 
of the exponential factor $e^{-m_Q^2/\Lambda}$ in Eq.~(\ref{d1}), are justified. It has been observed 
in the investigation of neutral meson mixing \cite{Li:2022jxc} that the optimal choice of $N$ for the 
polynomial expansion indeed increases with $\Lambda$ in the stability window.

\subsection{Charm Mass from the $Q\to d u\bar d$ Decay Width}

\begin{figure}
\begin{center}
\includegraphics[scale=0.35]{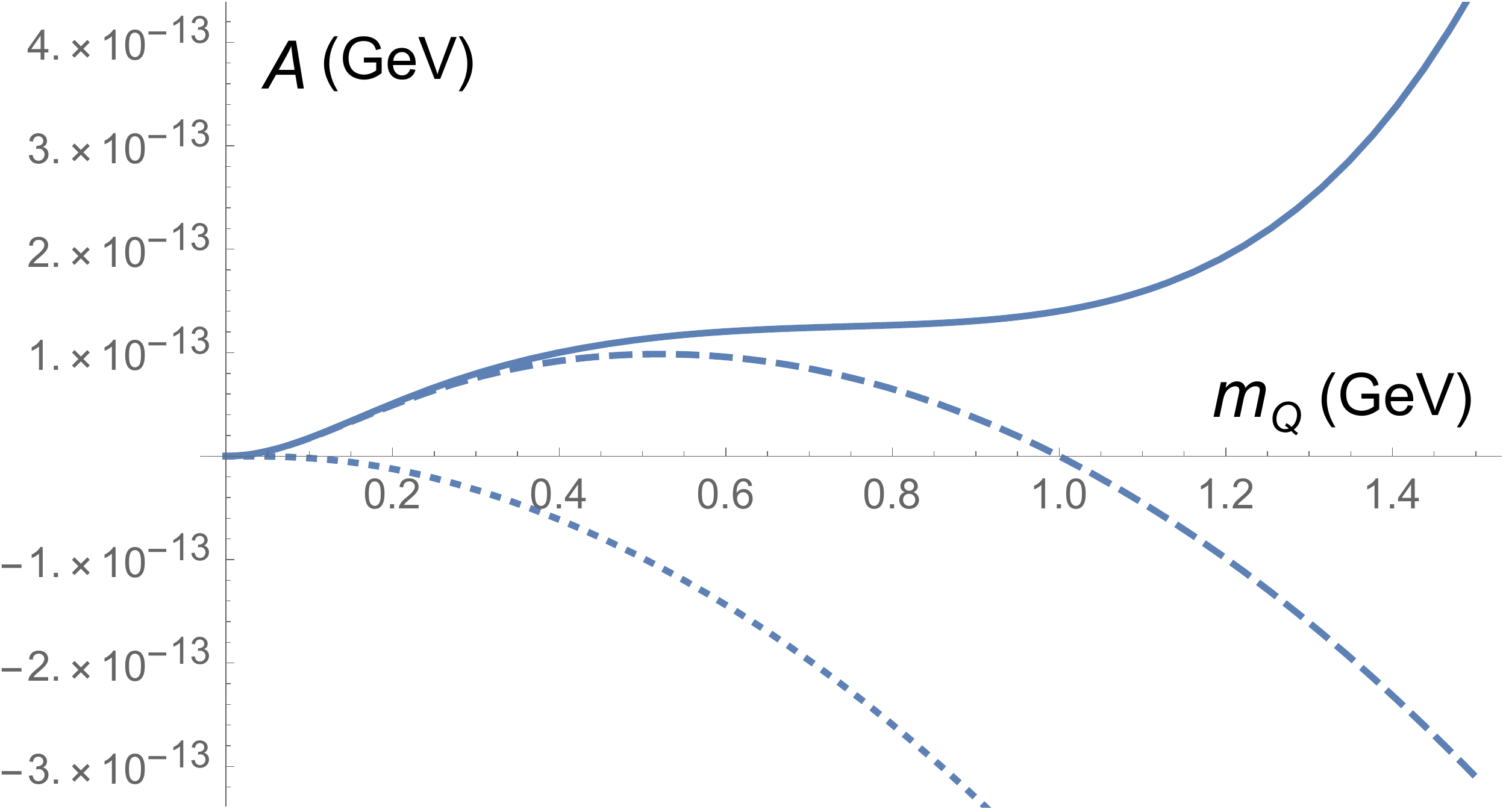}

\caption{\label{fig2}
The HQE hadronic decay width without the CKM factor,
$A(m_Q)\equiv \Gamma_h^{\rm HQE}(m_Q)/|V_{\rm CKM}|^2$, from the dimension-six contribution
(dotted line) only, from the sum of the dimension-six and -seven contributions (dashed line), and from 
the total contributions (solid line).}
\end{center}
\end{figure}

We deduce the charm quark mass from the dispersion relation for the hadronic decay 
$Q\to d u\bar d$, taking the Fermi constant $G_F=1.1663788\times 10^{-5}$ GeV$^{-2}$. Strictly 
speaking, the decay constant $f_{H_Q}$ depends on the fictitious meson mass $m_{H_Q}$. However, the 
decay constants of the physical pseudoscalar mesons do not vary much in the low mass region, ranging 
from $m_\pi\approx 0.14$ GeV to $m_{B_s}\approx 5.4$ GeV. Hence, we treat $f_{H_Q}$ 
as a constant, and set it to a typical value $f_{H_Q}=0.2$ GeV. The expansion for the heavy meson 
mass $m_{H_Q}=m_Q+\bar\Lambda$ in Eq.~(\ref{67}) is implemented. The contribution down by the power 
$\bar\Lambda/m_Q$ from the dimension-six term is then grouped into the dimension-seven one. The 
HQET parameters $\bar\Lambda$, $\mu_\pi^2$ and $\mu_G^2$ vary with $m_Q$ in principle. We also treat 
them as constants, and verify that outcomes have a weak dependence on them.
We take $\bar\Lambda=0.5$ GeV, $\mu_\pi^2=0.43$ GeV$^2$ and $\mu_G^2=0.38$ GeV$^2$, i.e.,
the central values of the ranges inferred from the evaluations of $B$ and $D$ meson lifetimes 
\cite{King:2021jsq,Lenz:2022rbq,Gratrex:2022xpm,Bernlochner:2022ucr,Kirk:2017juj}
\begin{eqnarray}
\bar\Lambda=0.5\pm 0.1\;{\rm GeV},\;\;\;\;
\mu_\pi^2=0.43\pm 0.24\;{\rm GeV}^2,\;\;\;\;
\mu_G^2=0.38\pm 0.07\;{\rm GeV}^2.
\end{eqnarray}
The renormalization group evolution of the Wilson coefficients $C_1(\mu)$ and $C_2(\mu)$ with 
$\mu=m_Q$ is taken into account to the leading logarithmic accuracy \cite{Buchalla:1995vs}. 
The fictitious quark mass can run to a very low scale in the $Q\to du\bar d$ case. To stabilize 
the running coupling constant which the Wilson coefficients depend on, we introduce an effective 
gluon mass $m_g$ into its argument:
\begin{eqnarray}
\alpha_s(\mu)=\frac{4\pi}{\beta_0\ln[(\mu^2+m_g^2)/\Lambda_{\rm QCD}^2]},
\end{eqnarray}
with the coefficient $\beta_0 = 11 - 2n_f/3$. We adopt the one-loop running with the QCD 
scale $\Lambda_{\rm QCD}=0.324$ GeV \cite{Zhong:2021epq} for the number of active quark flavors
$n_f=3$. The effective gluon mass has been estimated to be about $m_g\approx 0.4$ GeV 
\cite{Aguilar:2015bud,Gomez:2016xjz}. We choose $m_g=0.41$ GeV, and
the reason for this choice will be provided later.


The relative importance of various contributions to the HQE hadronic width in Eq.~(\ref{hq}) without 
the CKM factor, $A(m_Q)\equiv \Gamma_h^{\rm HQE}(m_Q)/|V_{\rm CKM}|^2$, is displayed in Fig.~\ref{fig2}. 
The dimension-six Pauli interference effect gives a negative contribution due 
to the combination of the Wilson coefficients $C_1^2+C_2^2+2N_cC_1C_2<0$ at a small scale 
\cite{Cheng:2018rkz}. The addition of the dimension-seven contribution turns the width positive 
up to the mass $m_Q\approx 1$ GeV. The decay width becomes positive 
definite, after all the contributions are included. The sum of the dimension-six 
and -seven contributions dominates the low mass region with $m_Q<0.4$ GeV. Therefore, 
we have the approximate $Q\to du\bar d$ HQE width at small $m_Q$,
\begin{eqnarray}
\Gamma_h^{\rm HQE}(m_Q)\approx \Gamma^{(6)+(7)}_{h}(m_Q) 
\propto m_Q^3-2\bar\Lambda m_Q^2.\label{hh}
\end{eqnarray}

\begin{figure}
\begin{center}
\includegraphics[scale=0.35]{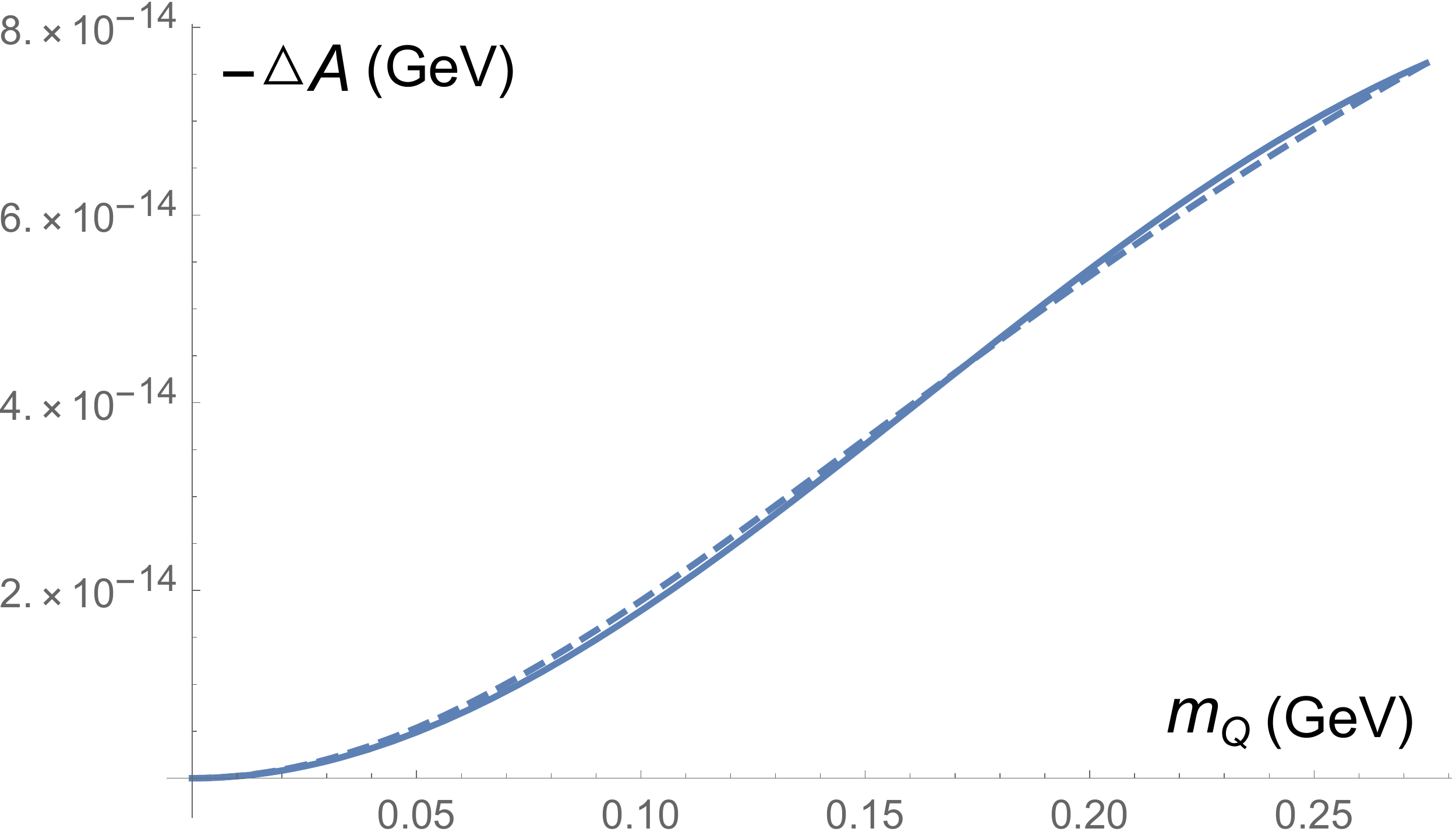}

\caption{\label{fig3}
Comparison of $\Delta\Gamma_{h}(m_Q)$ in Eq.~(\ref{so1}) from the fit (solid line) with 
$-\Gamma^{\rm HQE}_{h}(m_Q)$ (dashed line) in terms of the width without the CKM factor, 
$\Delta A(m_Q)\equiv \Delta\Gamma_h^{\rm HQE}(m_Q)/|V_{\rm CKM}|^2$.}
\end{center}
\end{figure}

Utilizing $J_\alpha(x)\sim (x/2)^\alpha/\Gamma(\alpha+1)$ as $x\to 0$, and comparing Eqs.~(\ref{do2}) 
and (\ref{hh}) at low $m_Q$, we identify the index $\alpha=1$ from the limiting behavior
$\Gamma_h^{\rm HQE}\propto m_Q^2$, and the ratio
\begin{eqnarray}
\frac{y_o}{y_e}=-\frac{1}{2\bar\Lambda}=-1\;{\rm GeV}^{-1}.\label{slo}
\end{eqnarray}
The coefficient $y_e$ is fixed
by the boundary condition $\Delta\Gamma_{h}(m_F)=-\Gamma^{\rm HQE}_{h}(m_F)$ at $m_Q=m_F$,
which leads the solution in Eq.~(\ref{do2}) to
\begin{eqnarray}
\Delta\Gamma_{h}(m_Q)=-\Gamma^{\rm HQE}_{h}(m_F)
\left(1-\frac{m_Q}{2\bar\Lambda}\right)\left(1-\frac{m_F}{2\bar\Lambda}\right)^{-1}
\frac{m_Q J_1(2\omega m_Q)}{m_F J_1(2\omega m_F)}.\label{so1}
\end{eqnarray}
The best fit of Eq~(\ref{so1}) to $-\Gamma^{\rm HQE}_{h}(m_Q)$ in the interval $(0,m_F)$, 
$m_F=m_{\pi^+}+m_{\pi^0}$ with the pion masses $m_{\pi^+}=0.140$ GeV and $m_{\pi^0}=0.135$ GeV, 
sets $\omega=\bar\omega=3.166$ GeV$^{-1}$. We contrast the fit result with 
$-\Gamma^{\rm HQE}_{h}(m_Q)$ in terms of the width without the CKM factor,
$\Delta A(m_Q)\equiv \Delta\Gamma_h^{\rm HQE}(m_Q)/|V_{\rm CKM}|^2$, in Fig.~\ref{fig3}. The 
excellent agreement confirms that the simple solution in Eq.~(\ref{do2}) works well, and that
other methods for determining $\bar\omega$, such as 
equating Eq.~(\ref{so1}) and $-\Gamma^{\rm HQE}_{h}(m_Q)$ at $m_Q=m_F/2$, yield
similar $\bar\omega$; this equality produces $\bar\omega=3.280$ GeV$^{-1}$, close to the one
from the best fit.

\begin{figure}
\begin{center}
\includegraphics[scale=0.35]{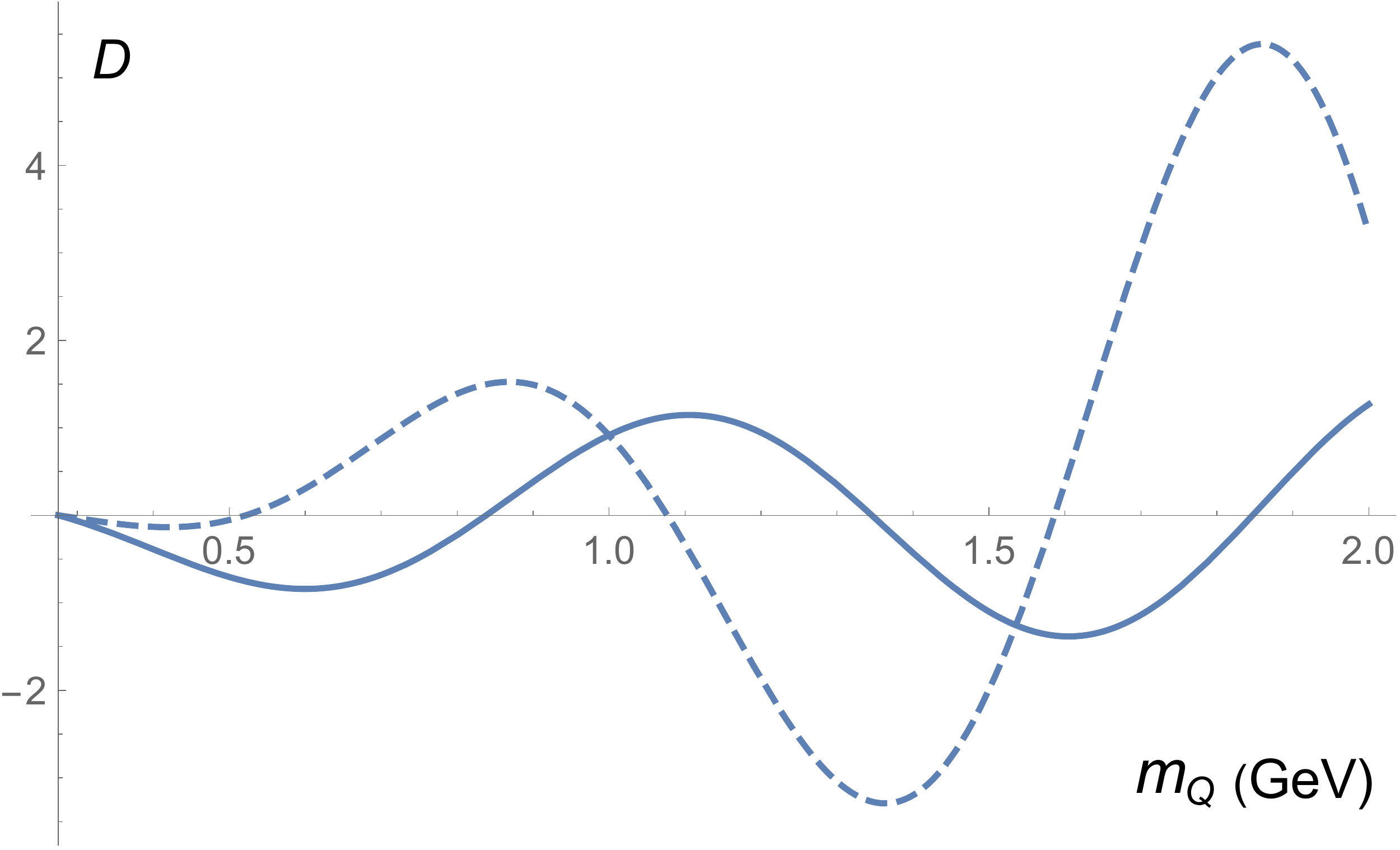}

\caption{\label{fig4}
Dependencies of the first derivative $D_1(m_Q)$ (in units of GeV, solid line) in Eq.~(\ref{der}) 
and the second derivative $D_2(m_Q)$ (in units of GeV$^2$, dashed line) in Eq.~(\ref{de2}) on $m_Q$.}
\end{center}
\end{figure}

As elaborated before, the charm quark mass takes the physical value $m_c$ that vanishes
the first derivative in Eq.~(\ref{dd1}), namely,
\begin{eqnarray}
D_1(m_Q)\equiv\frac{d}{d\omega}\frac{J_1(2\omega m_Q)}{J_1(2\omega m_F)}\Big|_{\omega=\bar\omega}=0,\label{der}
\end{eqnarray}
where the factors independent of $\omega$ in Eq.~(\ref{so1}) have been removed. At the same time, 
the second derivative 
\begin{eqnarray}
D_2(m_Q)\equiv \frac{d^2}{d\omega^2}\frac{J_1(2\omega m_Q)}{J_1(2\omega m_F)}
\Big|_{\omega=\bar\omega},\label{de2}
\end{eqnarray}
should be minimal as $m_Q=m_c$. Figure~\ref{fig4} presents the dependence of the first derivative in 
Eq.~(\ref{der}) on $m_Q$, which reveals several roots of $m_Q$ in the small $m_Q$ region. The first 
root located at $m_Q=m_F=0.275$ GeV, attributed to the boundary condition of $\Delta\Gamma_h(m_Q)$ at 
this $m_Q$, is trivial and bears no physical significance. The dashed curve for Eq.~(\ref{de2}) in 
Fig.~\ref{fig4} manifests a larger second derivative for, i.e., worse stability associated 
with a higher root, so a smaller root is preferred for $m_c$. It is known that a charm quark can 
decay into two strange quarks with the hadronic threshold $2m_K\approx 1$ GeV. The second root at 
$m_Q=0.84$ GeV seems too low to be physical. We thus select the third root at $m_Q=1.35$ GeV as the 
physical solution of the charm quark mass, which implies the $D^+$ meson mass 
$m_{D^+}=m_c+\bar\Lambda=1.85$ GeV in accordance with the measured value 1.870 GeV \cite{PDG}. 
Another support for this choice is that the corresponding decay width agrees with the data as shown 
later. Our solution of $m_c=1.35$ GeV is consistent with the $\overline{\rm MS}$, 
kinematic and pole masses of a charm quark, which are all equivalent at LO, and range between 
1.28 GeV to 1.49 GeV at one loop \cite{Gratrex:2022xpm}.

We examine the sensitivity of the extracted charm quark mass to the variation of the involved
inputs. The result has a weak dependence on the HQET parameters $\bar\Lambda$, $\mu_\pi$ and 
$\mu_G$. Taking the binding energy $\bar\Lambda$ as the representative example, we find that
the lower (upper) bound of the binding energy $\bar\Lambda=0.4$ GeV (0.6 GeV) 
generates the solution $m_c=1.37$ GeV (1.34 GeV). That is, 20\% change of $\bar\Lambda$ makes an 
impact of less than 2\% on $m_c$. The variation of the decay constant $f_{H_Q}$ has a similar effect: 
a smaller (larger) value $f_{H_Q}=0.16$ GeV (0.24 GeV) leads to $m_c=1.37$ GeV (1.33 GeV). 
Once the chiral symmetry is broken, up and down quarks should become massive too. 
We thus check the dependence on light quark masses by including the down quark mass $m_d=5$ MeV 
into the calculation, and assure that it has little influence on the outcome;
the charm quark mass just reduces from $m_c=1.35$ GeV to 1.34 GeV. The only 
parameter which $m_c$ is sensitive to is the effective gluon mass $m_g$: a smaller (larger)
$m_g=0.40$ GeV (0.42 GeV) gives $m_c=1.29$ GeV (1.40 GeV). This sensitivity is expected, for 
the hadronic threshold $m_F$, which the Wilson coefficients can evolve to, is quite low in this case. 
It is then understood why we chose $m_g=0.41$ GeV: the resultant $D^+$ meson mass $m_{D^+}=1.85$ GeV 
would be roughly equal to the observed one. We point out that the $Q\to d u\bar d$ decay is the only 
mode among those considered in the present work, whose analysis is sensitive to $m_g$. Once $m_g$ is 
set, it is employed in the investigations of the other modes, and the agreement of the solved 
fermion masses with the measured values will reinforce our claim that fermion masses in the SM 
are dynamically constrained.  

We also need to assess the theoretical uncertainty inherent in our formalism. Though the large $N$
approximations for establishing a solution are justified, it is not clear how large the highest 
degree $N'$ in Eq.~(\ref{d0}) is and how reliable the expression in terms of a single Bessel 
function in Eq.~(\ref{do2}) is. This uncertainty is reflected by that of the parameter $\bar\omega$ 
from matching the solution to the HQE input in the interval $(0,m_F)$---if a true solution 
was available, $\bar\omega$ should be determined unambiguously. Different ways of matching return 
different values of $\bar\omega$, as having been exemplified below Eq.~(\ref{so1}), and different 
results of $m_c$ accordingly. We estimate the error from this source 
by computing the squared deviation 
\begin{eqnarray}
\sigma\equiv\int_0^{m_F}\left[\Delta\Gamma_{h}(m_Q)\big|_{\omega=\bar\omega}+
\Gamma_h^{\rm HQE}(m_Q)\right]^2dm_Q.\label{squ}
\end{eqnarray}
A value of $\bar\omega$ is accepted, as $\sigma$ is lower than twice its minimum, which 
corresponds to the best fit. Given this prescription, it is straightforward to identify the 
allowed ranges 3.012 GeV$^{-1}<\bar\omega <3.306$ GeV$^{-1}$ and 1.29 GeV $<m_c <1.41$ GeV. Since 
the effective gluon mass will be fixed hereafter, the uncertainties surveyed above are dominated by 
the variation of $\bar\omega$, and sum to $m_c=1.35^{+0.07}_{-0.06}$ GeV.

\begin{figure}
\begin{center}
\includegraphics[scale=0.3]{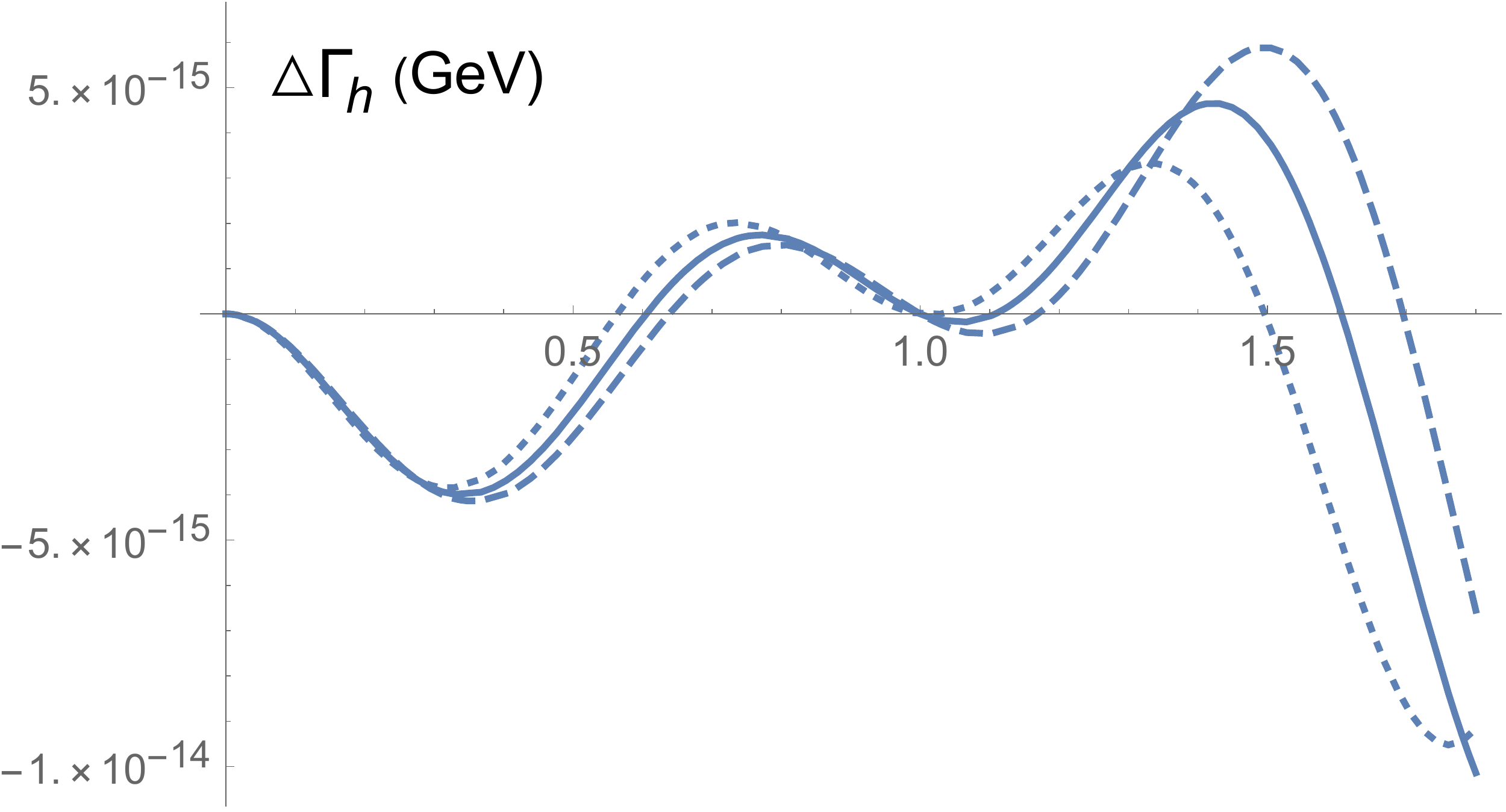}\hspace{1.0 cm} 
\includegraphics[scale=0.3]{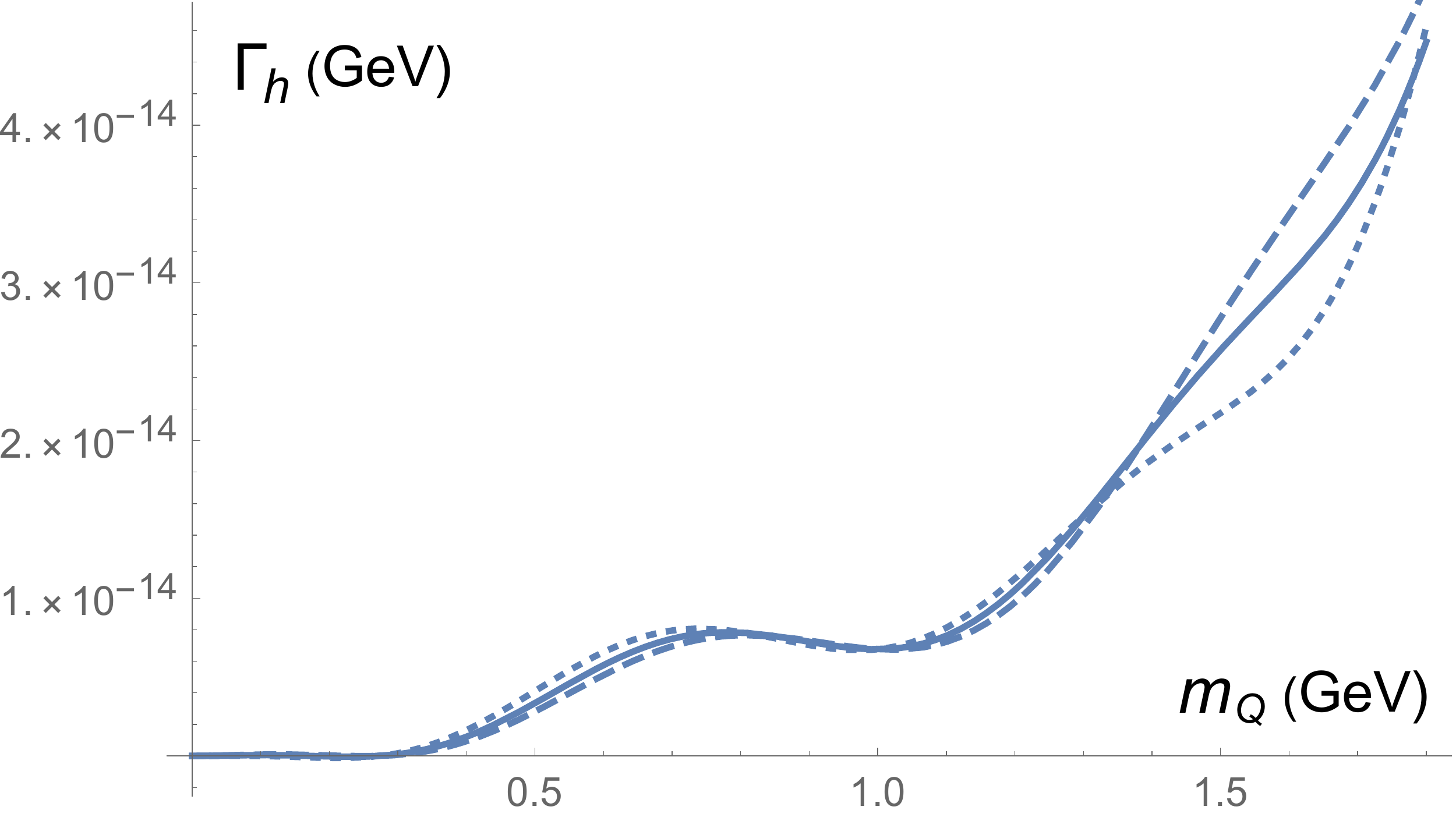}

(a) \hspace{7.0 cm} (b)
\caption{\label{fig5}
Dependencies of (a) the subtracted width $\Delta\Gamma_h(m_Q)$ and (b) the width $\Gamma_h(m_Q)$ 
of the $Q\to d u\bar d$ decay on $m_Q$ for $\omega=3.0$ GeV$^{-1}$ (dashed line), 
3.166 GeV$^{-1}$ (solid line) and 3.4 GeV$^{-1}$ (dotted line).}
\end{center}
\end{figure}

We then include the CKM factor $|V_{\rm CKM}|^2=|V_{cd}V_{ud}^*|^2$ with $V_{cd}=\lambda$ and
$V_{ud}=1-\lambda^2/2$, where the Wolfenstein parameter takes $\lambda=0.225$ \cite{PDG}.
The subtracted width $\Delta\Gamma_h(m_Q)$ and the width 
$\Gamma_h(m_Q)=\Delta\Gamma_h(m_Q)+\Gamma^{\rm HQE}_{h}(m_Q)$ 
of the $Q\to d u\bar d$ decay for three values of $\omega$ around $\bar\omega=3.166$ GeV$^{-1}$, 
i.e., for $\omega=3.0$ GeV$^{-1}$, 3.166 GeV$^{-1}$ and 3.4 GeV$^{-1}$ are plotted in 
Figs.~\ref{fig5}(a) and \ref{fig5}(b). Indeed all three curves cross each other in 
the compact regions located at $m_Q=0.84$ GeV and 1.35 GeV as expected, evincing the stability of 
the widths evaluated at these two values of $m_Q$ under the variation of $\omega$. The intersection 
area is smaller at $m_Q=0.84$ GeV than at $m_Q=1.35$ GeV, in alignment with the indication of the 
second derivative in Fig.~\ref{fig4}. However, we pick up the latter as the solution for the charm 
quark mass as explained before. It is mentioned that the similar phenomenon has occurred in the 
dispersive analysis of neutral meson mixing; the curves for the mixing parameters of a fictitious 
$D$ meson corresponding to various scales $\Lambda$ also cross each other in a compact region located 
at the $D$ meson mass. We clarify that the crossing at $m_Q=1$ GeV, also seen in 
Fig.~\ref{fig5}, is due to the vanishing of the factor $1-m_Q/(2\bar\Lambda)$ in Eq.~(\ref{so1}), 
which should not be mixed up with the roots of Eq.~(\ref{der}). The diminishing of the solved 
$\Gamma_h(m_Q)$ for small $m_Q$ up to $m_Q=m_F=0.275$ GeV in Fig.~\ref{fig5}(b) echoes the
almost exact cancellation between $\Delta\Gamma_h(m_Q)$ and $\Gamma^{\rm HQE}_{h}(m_Q)$ in the 
interval $(0,m_F)$, which has been illustrated in Fig.~\ref{fig3}.

The predicted widths at $m_Q=m_c=1.35$ GeV read 
$\Delta\Gamma_h(m_c)=4.09\times 10^{-15}$ GeV and $\Gamma_h(m_c)=1.79\times 10^{-14}$ GeV.
It signifies that the nonperturbative effect, originating from the introduction of the hadronic 
threshold, enhances the HQE result by about 30\%. The predicted hadronic decay width amounts
to the branching fraction $B(c\to du\bar d)=2.8\%$ for the inclusive Cabibbo-suppressed 
pionic modes, given the total decay width of the $D^+$ meson $\Gamma^{\rm tot}=6.37\times 10^{-13}$ 
GeV \cite{PDG}. This prediction is reasonable compared with the relevant data \cite{PDG}. The decay 
width $\Gamma_h(0.84\;{\rm GeV})=7.66\times 10^{-15}$ GeV at $m_Q=0.84$ GeV, amounting to the 
branching fraction 1.2\%, is apparently too low; the single channel $D^+\to 2\pi^+\pi^-\pi^0$ 
contributes the branching fraction about 1.2\% already according to \cite{PDG}.


\section{DISPERSIVE CONSTRAINTS ON OTHER FERMION MASSES}

We extend the formalism developed in the previous section to the constraints on the other fermion 
masses, including the strange quark mass $m_s$ from the $Q\to su\bar d$ decay,
the bottom quark mass $m_b$ from the $Q\to c\bar ud$ decay, the muon mass $m_\mu$ from
the $Q\to d\mu^+ \nu_\mu$ decay and the $\tau$ lepton mass $m_\tau$ from the 
$Q\to u\tau^- \bar\nu_\tau$ decay. For the constraints on $m_s$, $m_\mu$ and $m_\tau$, we
rely on the proposition that dispersive analyses of different decay channels of a fictitious 
heavy quark should conclude the same heavy quark mass.

\subsection{The $Q\to su\bar d$ and $Q\to c\bar ud$ Decays}

We first derive a solution to the dispersion relation obeyed by hadronic decay widths for 
massive final states, e.g., with nonvanishing $m_q$ in Eq.~(\ref{67}). Similarly, we decompose 
the HQE input into the sum of the even and odd pieces
$\Gamma_h^{\rm HQE}(m)=\Gamma_e^{\rm HQE}(m)+\Gamma_o^{\rm HQE}(m)$, 
and the unknown width into $\Gamma_h(m)=\Gamma_e(m)+\Gamma_o(m)$.  
Note that there exists an additional pole at $m_Q=0$ in the HQE result as shown in Eq.~(\ref{67}),
\begin{eqnarray}
\Gamma^{(6)+(7)}_{h}(m_Q) \propto \frac{(m_Q^2-m_q^2)^2}{m_Q}
-4\bar\Lambda (m_Q^2-m_q^2),
\label{mh}
\end{eqnarray}
where the expansion of the factor $1+m_q^2/m_Q^2\approx 2+ O((m_Q^2-m_q^2)/m_Q^2)$ in the limit 
$m_Q\to m_q$ has been applied to the second term. The single pole in the dimension-six term requests 
a slightly different handling as elaborated below. The dispersion relation for the even piece is 
similar to Eq.~(\ref{ge}), except that the lower bound of $m^2$ is replaced by the quark-level 
threshold $m_q^2$,
\begin{eqnarray}
\int_{m_q^2}^\infty\frac{\Delta\Gamma_e(m)}{m_Q^2-m^{2}}dm^2=0.\label{15}
\end{eqnarray}
The unknown function $\Delta\Gamma_e(m)$ is fixed to $-\Gamma_e^{\rm HQE}(m)$ in the 
interval $(m_q,m_F)$ of $m$ with the hadronic threshold $m_F$.



For the odd piece, we consider the contour integration of the correlator $m^2\Pi(m)/(m^2-m_q^2)$, 
for which the high-mass behavior of $\Pi(m)$ is not altered, and the $m=0$ pole has been 
removed, so that Eq.~(\ref{con}) holds. The additional poles at $m=\pm m_q$ are introduced, but 
their contribution $m_qM(m_q)/m_Q$ to the left-hand side of Eq.~(\ref{ij}) is much smaller than 
$M(m_Q)$ from the pole $m=m_Q$ at large $m_Q$. A similar contribution from the poles at 
$m=\pm m_q$ also exists on the left-hand side of Eq.~(\ref{ope}). We still equate the left-hand sides 
of Eqs.~(\ref{ij}) and (\ref{ope}), and this equality is justified, as long as the odd piece 
$m^2\Gamma_o(m)/(m^2-m_q^2)$ can be solved from the dispersion relation
\begin{eqnarray}
& &\int_{m_F}^R\frac{m^2\Gamma_o(m)}{(m^2-m_q^2)(m_Q-m)}dm-
\int^{-m_F}_{-R}\frac{m^2\Gamma_o(m)}{(m^2-m_q^2)(m_Q-m)}dm\nonumber\\
&=&\int_{m_q}^R\frac{m^2\Gamma_o^{\rm HQE}(m)}{(m^2-m_q^2)(m_Q-m)}dm-
\int^{-m_q}_{-R}\frac{m^2\Gamma_o^{\rm HQE}(m)}{(m^2-m_q^2)(m_Q-m)}dm.\label{mas}
\end{eqnarray}
Employing the variable change $m\to -m$ for the second integrals on both sides
and moving the integrands on the right-hand side to the left-hand side, we get
\begin{eqnarray}
\int_{m_q^2}^\infty\frac{m\Delta\Gamma_o(m)}{(m^2-m_q^2)(m_Q^2-m^{2})}dm^2=0,\label{19}
\end{eqnarray}
where the upper bound of the integration variable $m^2$ has been extended to infinity. The unknown 
function $\Delta\Gamma_o(m)$ is fixed to $-\Gamma_o^{\rm HQE}(m)$ in the interval 
$(m_q,m_F)$.


The steps from Eq.~(\ref{i2}) to Eq.~(\ref{o2}) can be carried out for the current case with massive 
final states straightforwardly. The only modification resides in the variable changes
$m_Q^2-m_q^2=u\Lambda$ and $m^2-m_q^2=v\Lambda$. We then construct the solutions
\begin{eqnarray}
\Delta \Gamma_e(m)&\approx&
y_e\left(\omega \sqrt{m^2-m_q^2}\right)^{\alpha} J_\alpha\left(2\omega\sqrt{m^2-m_q^2}\right),
\label{md2}\\
\frac{m\Delta \Gamma_o(m)}{m^2-m_q^2}&\approx&
y_o\left(\omega \sqrt{m^2-m_q^2}\right)^{\alpha} J_\alpha\left(2\omega\sqrt{m^2-m_q^2}\right).
\label{mo2}
\end{eqnarray} 
A solution of the hadronic decay width is written as
\begin{eqnarray}
\Delta \Gamma_h(m_Q)=\Delta \Gamma_e(m_Q)+\Delta \Gamma_o(m_Q)\approx
y_e\left(1+\frac{y_o}{y_e} \frac{m_Q^2-m_q^2}{m_Q}\right)
\left(\omega \sqrt{m_Q^2-m_q^2}\right)^{\alpha} 
J_\alpha\left(2\omega\sqrt{m_Q^2-m_q^2}\right),\label{eo2}
\end{eqnarray}
which reduces to Eq.~(\ref{do2}) as $m_q=0$ obviously.

We apply Eq.~(\ref{eo2}) to the channel $Q\to su\bar d$ with the thresholds $m_q=m_s$ and 
$m_F=m_{\pi^+}+m_{\bar K^0}$. Comparing Eqs.~(\ref{eo2}) and (\ref{mh}) 
in the limit $m_Q\to m_q$, we identify the index $\alpha=1$ and the ratio
\begin{eqnarray}
\frac{y_o}{y_e}=-\frac{1}{4\Lambda}=-0.5\;{\rm GeV}^{-1}.
\end{eqnarray}
The boundary condition $\Delta\Gamma_{h}(m_F)=-\Gamma^{\rm HQE}_{h}(m_F)$ at $m_Q=m_F$ leads 
Eq.~(\ref{eo2}) to
\begin{eqnarray}
\Delta\Gamma_{h}(m_Q)=-\Gamma^{\rm HQE}_{h}(m_F)
\left(1-\frac{m_Q^2-m_s^2}{4\bar\Lambda m_Q}\right)
\left(1-\frac{m_F^2-m_s^2}{4\bar\Lambda m_Q}\right)^{-1}
\frac{\sqrt{m_Q^2-m_s^2} J_1\left(2\omega \sqrt{m_Q^2-m_s^2}\right)}
{\sqrt{m_F^2-m_s^2}J_1\left(2\omega \sqrt{m_F^2-m_s^2}\right)}.\label{dc2}
\end{eqnarray}
With the same binding energy $\bar\Lambda=0.5$ GeV and effective gluon mass $m_g=0.41$ GeV,
and the meson masses $m_{\pi^+}=0.140$ GeV and $m_{\bar K^0}=0.498$ GeV, we obtain 
$\bar\omega=1.995$ GeV$^{-1}$, $2.080$ GeV$^{-1}$ and $2.177$ GeV$^{-1}$ for the strange quark 
mass $m_s=0.01$ GeV, 0.12 GeV and 0.30 GeV, respectively, from the best fit of Eq.~(\ref{dc2})
to $-\Gamma^{\rm HQE}_{h}(m_Q)$ in the interval $(m_s,m_F)$. 

\begin{figure}
\begin{center}
\includegraphics[scale=0.35]{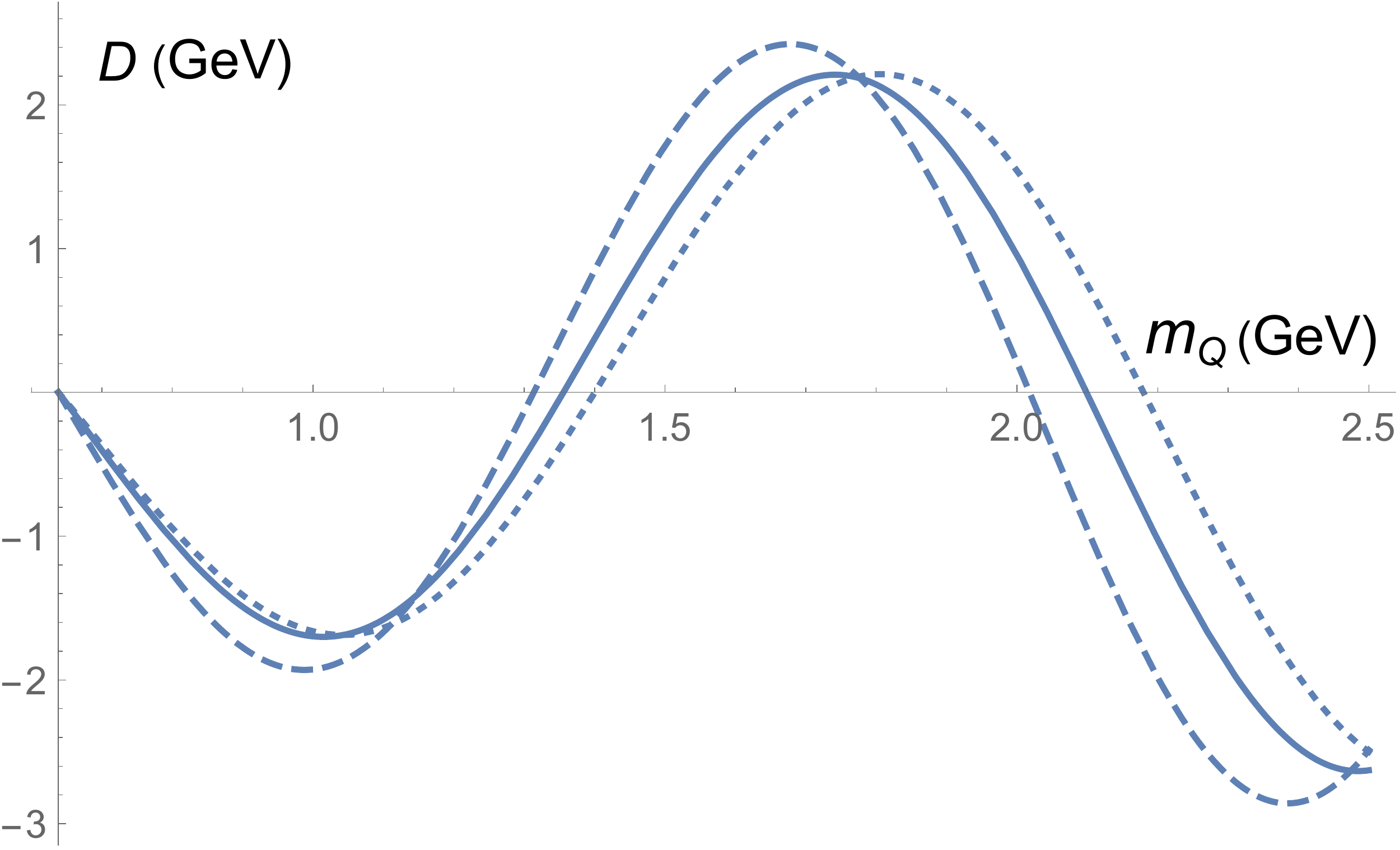}
\caption{\label{fig6} 
Dependencies of $D(m_Q)$ in Eq.~(\ref{der2}) on $m_Q$ for $m_s=0.01$ GeV
(dotted line), 0.12 GeV (solid line) and 0.30 GeV (dashed line).}
\end{center}
\end{figure}

The stability of the solved decay width under the variation of $\omega$ demands 
the vanishing of the derivative $d\Delta\Gamma_{h}(m_Q)/d\omega$ at $\omega=\bar\omega$, i.e.,
\begin{eqnarray}
D(m_Q)\equiv \frac{d}{d\omega}\frac{J_1\left(2\omega \sqrt{m_Q^2-m_s^2}\right)}
{J_1\left(2\omega \sqrt{m_F^2-m_s^2}\right)}\Big|_{\omega=\bar\omega}=0.\label{der2}
\end{eqnarray}
The dependencies of the above derivative on $m_Q$ for the three values of $\bar\omega$ are 
exhibited in Fig.~\ref{fig6}, where the first roots located at $m_Q=m_F$ have no physical 
significance, because they arise from the boundary condition. We read off the second roots 
$m_Q=1.40$ GeV, 1.35 GeV and 1.31 GeV for $\bar\omega=1.995$ GeV$^{-1}$, $2.080$ GeV$^{-1}$ and 
$2.177$ GeV$^{-1}$ ($m_s=0.01$ GeV, 0.12 GeV and 0.30 GeV), respectively, at which the solutions 
possess the maximal stability as argued in the previous section. Those roots at higher $m_Q$ with 
worse stability are not selected. It is encouraging to find that the strange quark must have a mass 
around $m_s=0.12$ GeV in order to maintain the charm quark mass $m_c=1.35$ GeV, the same as from 
the $Q\to d u\bar d$ analysis. A lower (higher) $m_s$ would give rise to a higher (lower) $m_c$. 
We have also checked that it is impossible to reproduce the charm quark
mass without the spectator contribution in Eq.~(\ref{mh}); $m_c$ is always higher 
than 3.0 GeV for any $m_s<m_K$, once the spectator contribution is switched off.
That is, the higher-power effect is necessary for establishing the physical solution.

The result $m_s=0.12$ GeV is consistent with the one derived from the known kaon mass $m_K$ in QCD 
sum rules at the scale $\mu=m_c=1.35$ GeV \cite{Dominguez:2007my}. The above observation confirms 
the nontrivial correlation among the masses $m_s$, $m_\pi$, $m_K$ 
and $m_c$, which characterize strong and weak dynamics in the SM, though their exact values may 
suffer potential theoretical uncertainties from, say, higher-order corrections to the HQE input.
It is verified that the extractions from the $Q\to su\bar d$ decay width are 
less sensitive to the variation of the effective gluon mass $m_g$. It can take a value as low (high) 
as 0.35 GeV (0.46 GeV) to decrease (increase) $m_c$ to 1.30 GeV (1.40 GeV) for $m_s=0.12$ GeV. In the 
previous $Q\to du\bar d$ case, $m_g$ takes 0.40 GeV (0.42 GeV) to make the same amount of 
impact to the solution of $m_c$. The influence from the other parameters, like the HQET ones, is 
also milder. To estimate the uncertainty from the variation of $\bar\omega$, we search for its value  
for each $m_s$, which generates $m_c=1.35$ GeV from Eq.~(\ref{der2}). It is then examined 
whether the squared deviation for this set of $\bar\omega$ and $m_s$, defined similarly to 
Eq.~(\ref{squ}) but with the integration interval $(m_s,m_F)$, is below twice its minimum.
If it is, the considered $m_s$ value is accepted. Iterating the procedure for various
$m_s$, we can acquire the allowed range of $m_s$ in principle. It turns out that the resultant range 
$m_s<0.4$ GeV is very wide. On one hand, the above error estimate may be too conservative to achieve
effective bounds in this case. On the other hand, it discloses the worse quality of the solution 
compared to that for the $Q\to d u\bar d$ decay. The worse quality could be attributed to the more 
complicated functional form of the HQE input for massive final states than for massless 
ones, such that the simple solution in Eq.~(\ref{dc2}) is less accurate than in
Eq.~(\ref{do2}). Therefore, we do not attach errors to the determination of $m_s$. 


\begin{figure}
\begin{center}
\includegraphics[scale=0.3]{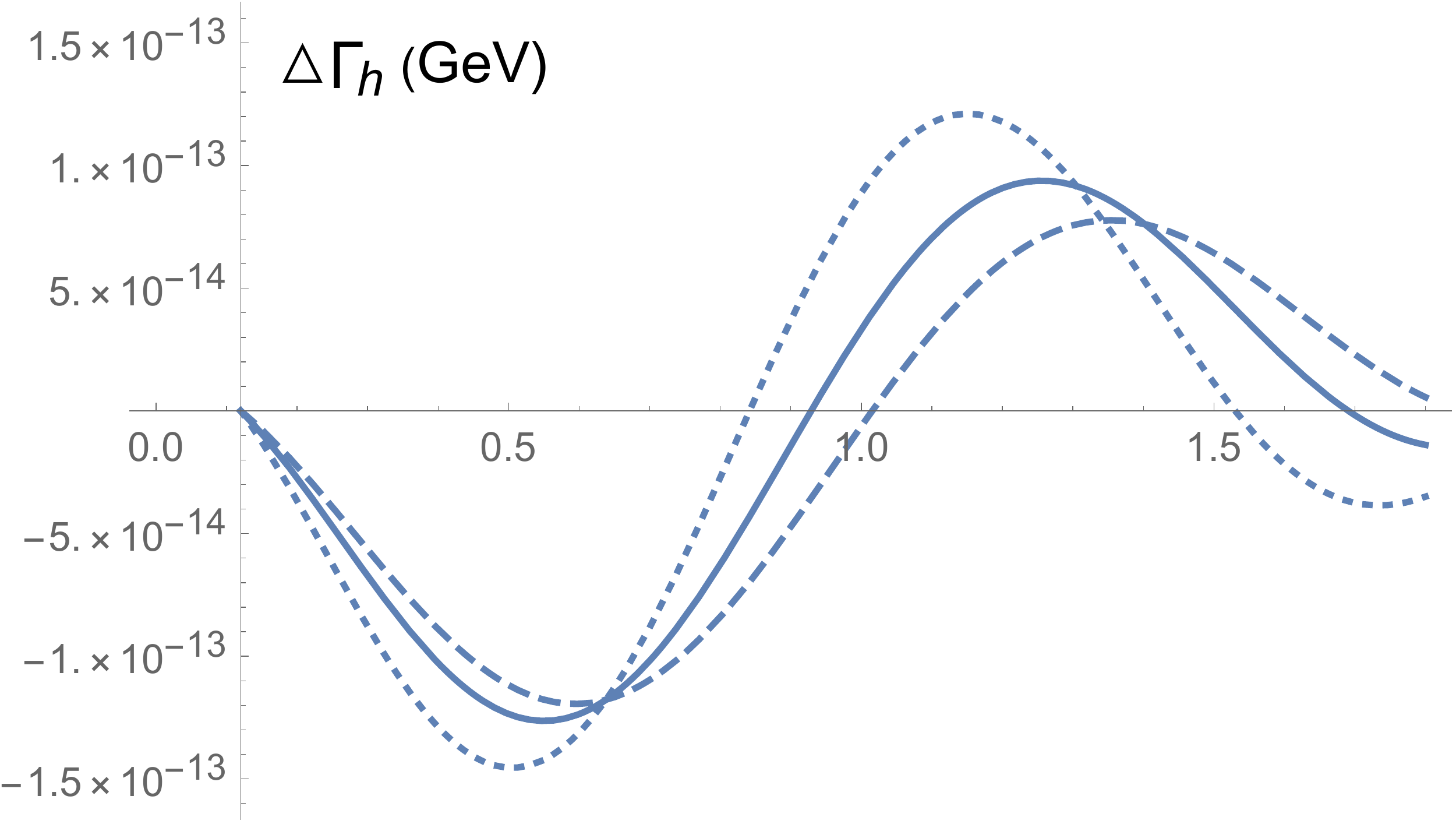}\hspace{1.0 cm} 
\includegraphics[scale=0.3]{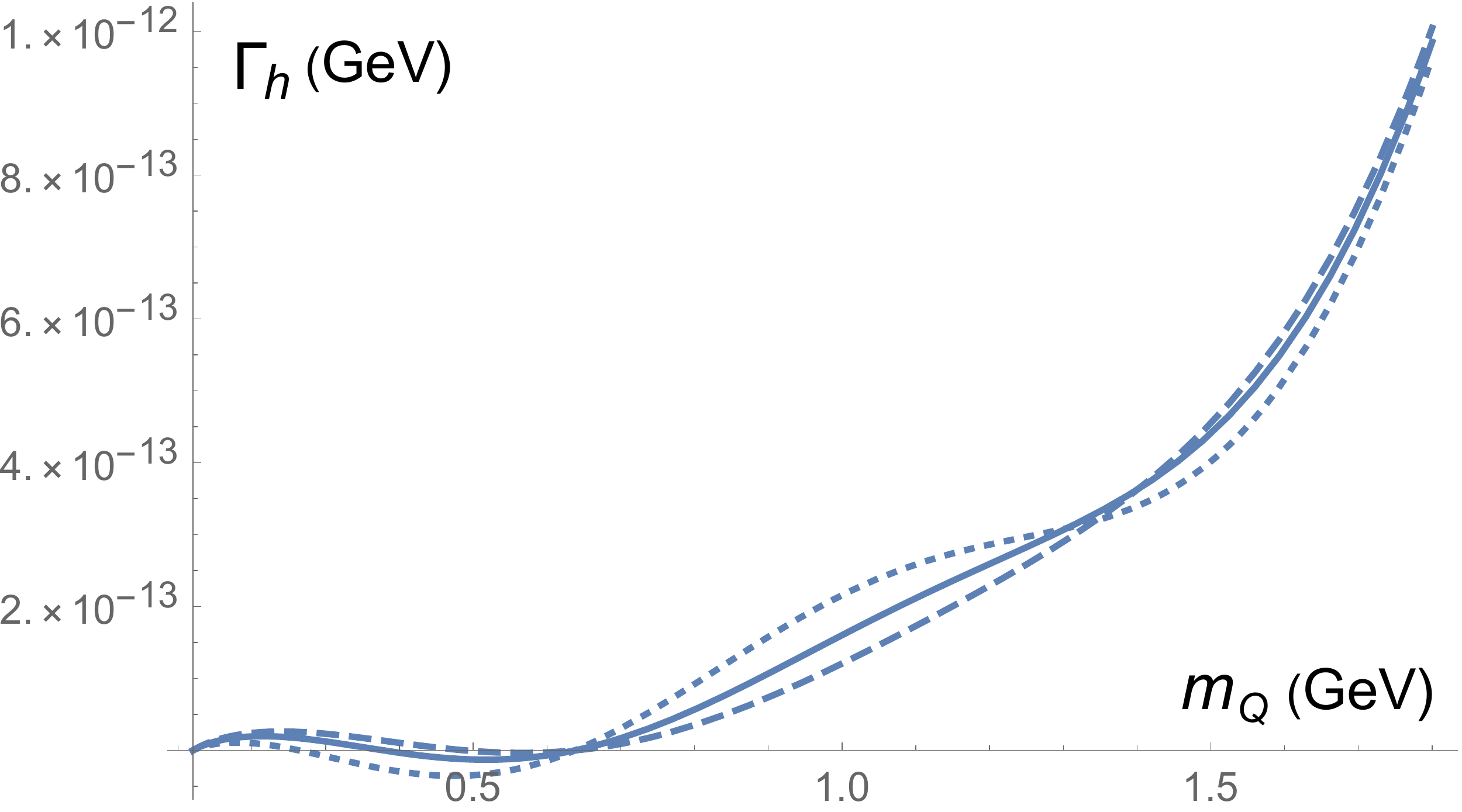}

(a) \hspace{7.0 cm} (b)
\caption{\label{fig7} 
Dependencies of (a) the subtracted width $\Delta\Gamma_h(m_Q)$ and (b) the width $\Gamma_h(m_Q)$ 
of the $Q\to s u\bar d$ decay on $m_Q$ for $\omega=1.9$ GeV$^{-1}$ (dashed line), 
2.080 GeV$^{-1}$ (solid line) and 2.3 GeV$^{-1}$ (dotted line).}
\end{center}
\end{figure}

The $m_Q$ dependencies of the subtracted width $\Delta \Gamma_h(m_Q)$ in Eq.~(\ref{dc2}) and 
the width $\Gamma_h(m_Q)$ of the $Q\to s u\bar d$ decay for three values of $\omega$ around 
$\bar\omega$ from the best fit, i.e., $\omega=1.9$ GeV$^{-1}$, 2.080 GeV$^{-1}$ and 2.3 GeV$^{-1}$,
are plotted in Figs.~\ref{fig7}(a) and \ref{fig7}(b), respectively, where the CKM factor 
$|V_{\rm CKM}|^2=|V_{cs}V_{ud}^*|^2$ has been included with $V_{cs}=1-\lambda^2/2$. The small
$\Gamma_h(m_Q)$ in the interval $(m_s,m_F)$ of $m_Q$ with $m_s=0.12$ GeV and 
$m_F=0.638$ GeV in Fig.~\ref{fig7}(b) reflects the satisfactory match between $\Delta\Gamma_h(m_Q)$ 
and $-\Gamma^{\rm HQE}_{h}(m_Q)$. The three curves cross each other more tightly 
at $m_Q=1.35$ GeV, implying the stability of the widths at this $m_Q$  
under the variation of $\omega$. The crossing occurring at $m_Q=m_F=0.638$ GeV, due to the boundary 
condition, bears no physical significance. The predicted widths $\Delta\Gamma_h(m_c)=8.61\times 10^{-14}$ 
GeV and $\Gamma_h(m_c)=3.33\times 10^{-13}$ GeV at $m_c=1.35$ GeV can be read off easily.
It indicates that the nonperturbative effect, originating from the introduction of the 
hadronic threshold, enhances the HQE result by about 35\%. The above decay width 
amounts to the branching fraction $B(c\to su\bar d)=52.3\%$ for the inclusive Cabibbo-favored
modes. Since the branching fraction of the semileptonic $D^+$ meson decays is about 
34\% \cite{PDG}, and the Cabibbo-favored modes dominate the hadronic channel, our prediction
is reasonable. Thanks to the nonperturbative enhancement observed in our formalism, the charmed meson 
lifetimes can be accommodated without resorting to a large fitted charm quark mass 1.56 GeV 
\cite{Cheng:2018rkz}.

\begin{figure}
\begin{center}
\includegraphics[scale=0.35]{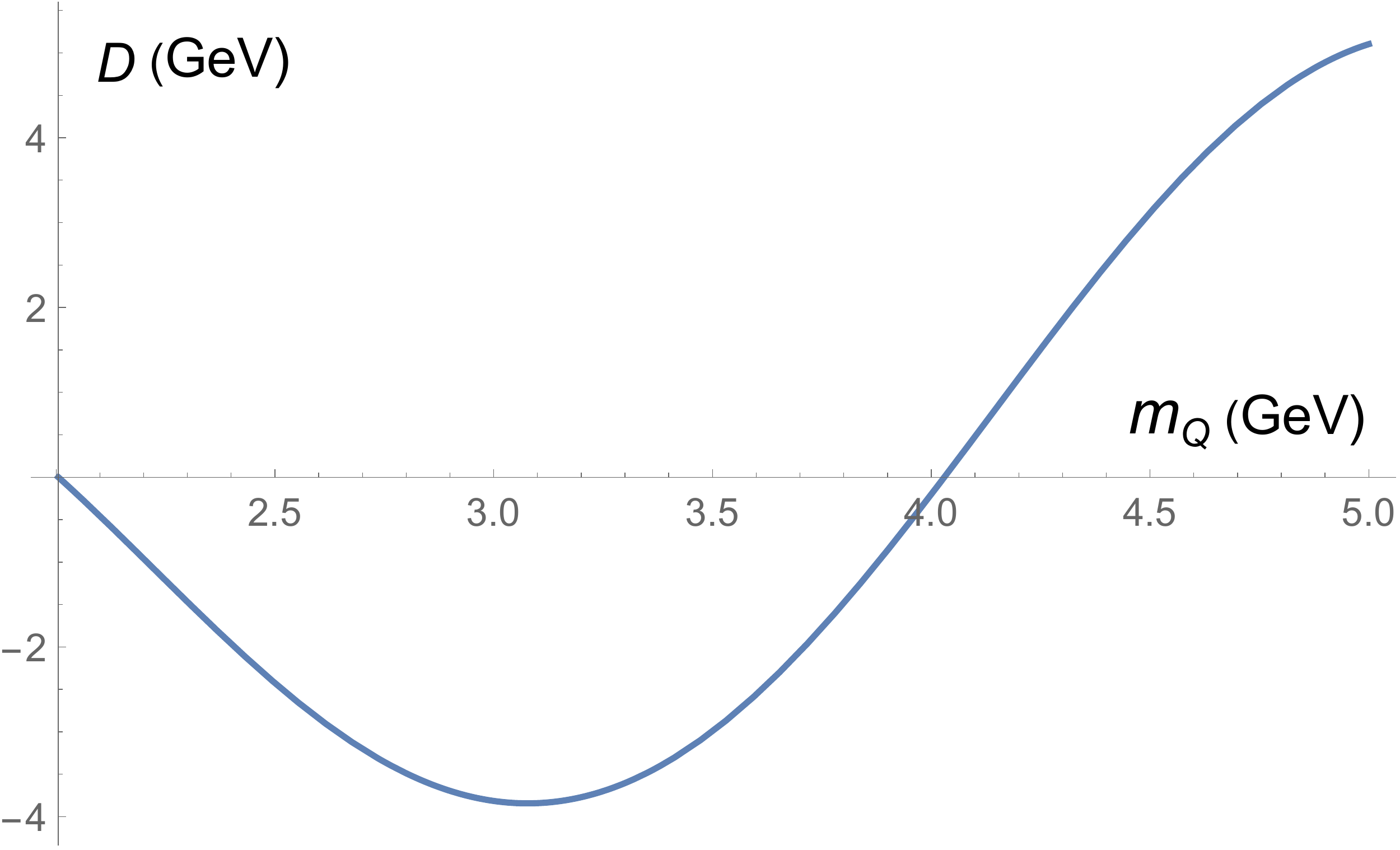}
\caption{\label{fig8} 
Dependencies of $D(m_Q)$ in Eq.~(\ref{der2}) on $m_Q$ for the $Q\to c u\bar d$ decay.}
\end{center}
\end{figure}

We turn to the analysis of the $Q\to c\bar ud$ decay width, simply 
substituting $m_c$ for the final-state quark mass $m_s$ and $m_F=m_{\pi^+}+m_{D^0}$ for the threshold 
$m_F=m_{\pi^+}+m_{\bar K^0}$. Another modification occurs in the choice of the QCD scale 
$\Lambda_{\rm QCD}=0.286$ GeV for the active quark number $n_f=4$ \cite{Zhong:2021epq}.
Taking the same binding energy $\bar\Lambda=0.5$ GeV and effective gluon mass 
$m_g=0.41$ GeV, $m_c=1.35$ GeV extracted previously and $m_{D^0}=1.865$ GeV \cite{PDG}, we derive 
$\bar\omega=0.712$ GeV$^{-1}$ from the best fit of the solution in Eq.~(\ref{dc2}) to the constraint 
$-\Gamma^{\rm HQE}_{h}(m_Q)$ in the interval $(m_c,m_F)$. Figure~\ref{fig8} shows the dependence of 
the derivative in Eq.~(\ref{der2}) on $m_Q$, and the second root located at $m_Q=4.03$ GeV corresponds 
to the physical bottom quark mass $m_b$, close to the running quark mass $\bar m_b(\bar m_b)=4.248$ 
GeV \cite{Cheng:2018rkz}. We stress that the formula for $\Gamma^{\rm HQE}_h(m_Q)$ does not carry 
any a priori information on a bottom quark, the emergence of $m_b$ is nontrivial, and the dispersion 
relation correlates the masses $m_c$, $m_D$ and $m_b$. This correlation is solid in the sense 
that it is not affected by the variation of $m_g$, because of the 
sizable threshold $m_F$. For instance, $\pm 25\%$ change of $m_g$ induces only $\pm 0.5\%$ change 
of $m_b$. We simply concentrate on the theoretical error inherent in our formalism. The condition 
that the squared deviation, defined similarly to Eq.~(\ref{squ}), does not exceed twice its 
minimum leads to the allowed range 0.706 GeV$^{-1}<\bar\omega <0.718$ GeV$^{-1}$. We thus get
the range of the bottom quark mass 4.01 GeV $<m_b <4.06$ GeV, and conclude  
$m_b=4.03^{+0.03}_{-0.02}$ GeV.

One may wonder that the predicted $B$ meson mass $m_B=m_b+\bar\Lambda=4.53$ is lower than 
the measured value 5.279 GeV \cite{PDG}. It may be due to the same HQET input $\bar\Lambda=0.5$ 
GeV assumed in the analyses of the $D$ and $B$ meson decays. In fact, the binding 
energy $\bar\Lambda$ can be $m_Q$-dependent as mentioned before. If $\bar\Lambda(m_Q)$ increases with 
$m_Q$, we will be allowed to test a larger value, say, $\bar\Lambda=0.6$ GeV for the $Q\to c\bar ud$ 
decay, which yields $m_b=4.42$ GeV. It lifts the predicted $B$ meson mass up to 5.02 GeV, approaching  
the observed one.

\begin{figure}
\begin{center}
\includegraphics[scale=0.3]{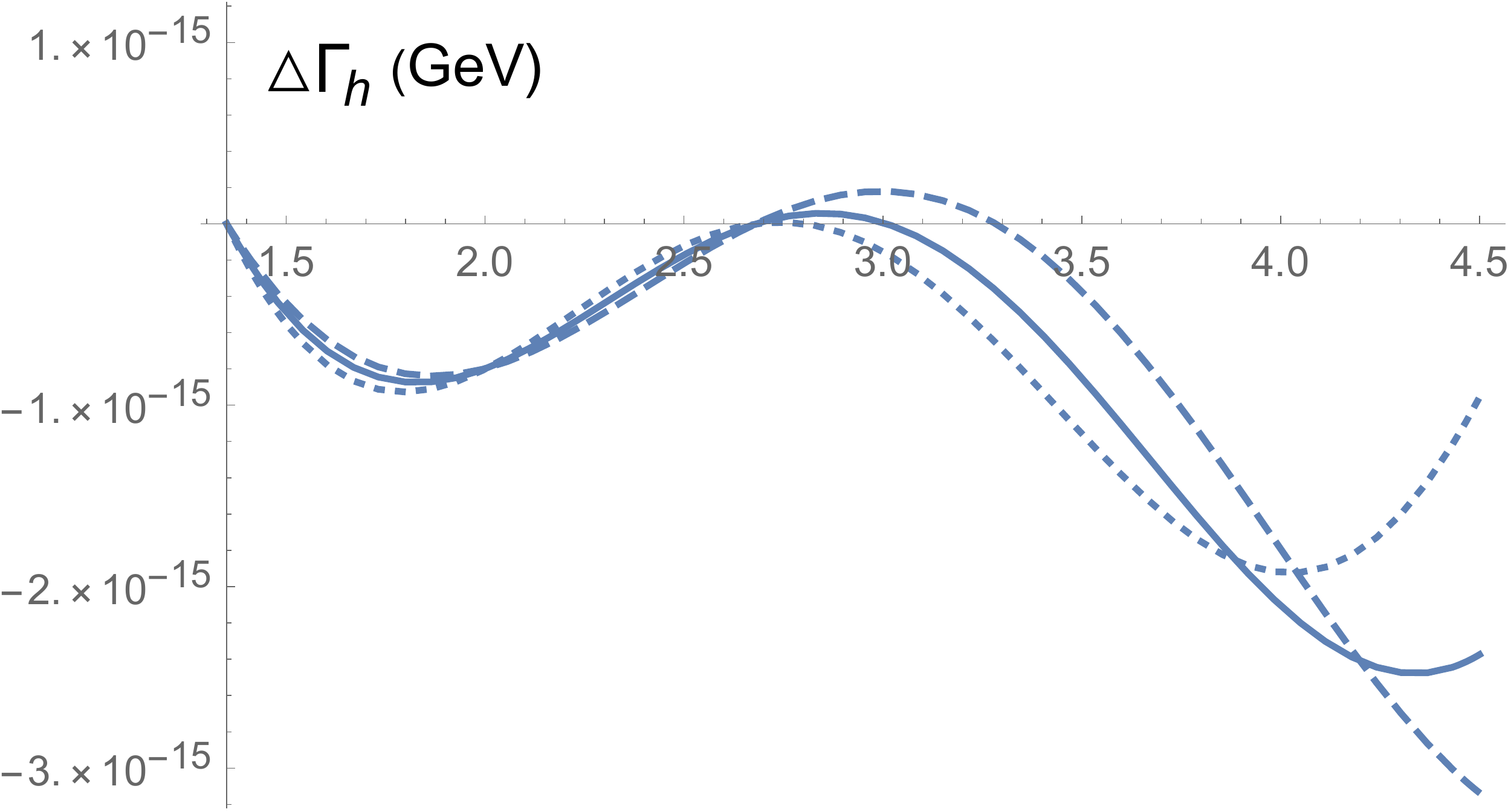}\hspace{1.0 cm} 
\includegraphics[scale=0.3]{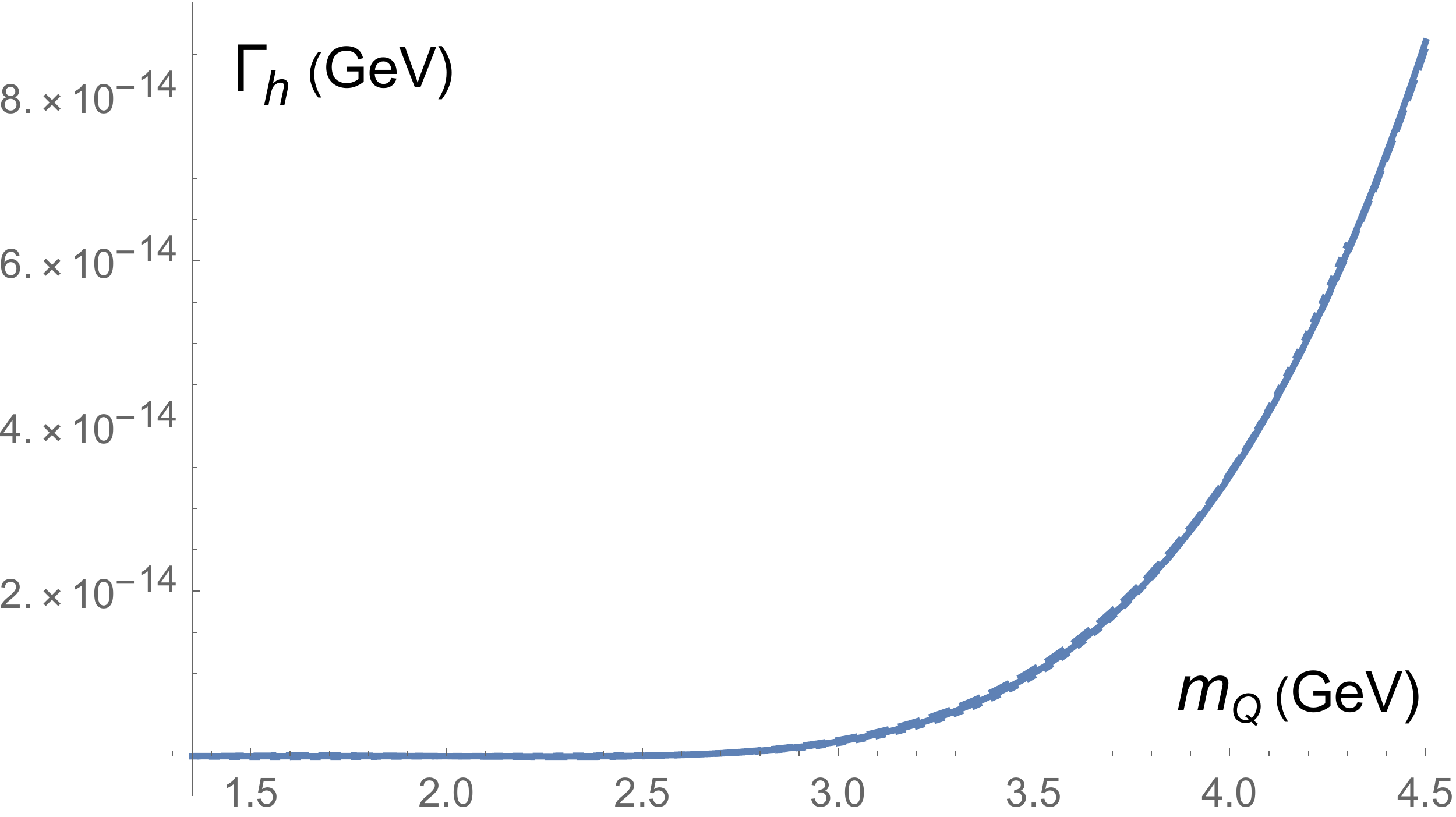}

(a) \hspace{7.0 cm} (b)
\caption{\label{fig9} 
Dependencies of (a) the subtracted width $\Delta\Gamma_h(m_Q)$ and (b) the width $\Gamma_h(m_Q)$ 
of the $Q\to c u\bar d$ decay on $m_Q$ for $\omega=0.64$ GeV$^{-1}$ (dashed line), 
0.712 GeV$^{-1}$ (solid line) and 0.78 GeV$^{-1}$ (dotted line).}
\end{center}
\end{figure}

We present the dependencies of the subtracted width $\Delta \Gamma_h(m_Q)$ and the width 
$\Gamma_h(m_Q)$ of the $Q\to c\bar ud$ decay with three different $\omega$ values in Figs.~\ref{fig9}(a) 
and \ref{fig9}(b), respectively, where the CKM factor $|V_{\rm CKM}|^2=|V_{cb}V_{ud}^*|^2$ has been 
included with $V_{cb}=A\lambda^2/2$, $A= 0.826$ being one of the Wolfenstein parameters \cite{PDG}. 
It is noticed that the three curves corresponding to $\omega=0.64$ GeV$^{-1}$, 0.712 GeV$^{-1}$ and 
0.78 GeV$^{-1}$ cross each other in the small region located at $m_Q\approx 4$ GeV in 
Fig.~\ref{fig9}(a). The crossings at $m_Q=m_F=2.005$ GeV resulting from the boundary condition,
and at $m_Q=2.680$ which vanishes the factor $1-(m_Q^2-m_c^2)/(4\bar\Lambda m_Q)$ in Eq.~(\ref{dc2}), 
have no physical significance. The three curves are on top of one another in Fig.~\ref{fig9}(b), and 
their intersection is barely seen, for the magnitude of $\Delta \Gamma_h(m_Q)$ is much lower than 
$\Gamma_h(m_Q)$. The smallness of $\Delta \Gamma_h(m_Q)$ reveals a minor nonperturbative 
contribution to $B$ meson decays introduced by the hadronic threshold $m_F$, and explains the 
reliability of HQE for the evaluation of $B$ meson lifetimes. The solved $\Gamma_h(m_Q)$ remains 
vanishing in the interval $(m_c,m_F)$ of $m_Q$ (in fact, up to $m_Q=2.5$ GeV) in Fig.~\ref{fig9}(b), 
manifesting the perfect match between $\Delta\Gamma_h(m_Q)$ and $-\Gamma^{\rm HQE}_{h}(m_Q)$.
We read off the widths $\Delta \Gamma_h(m_b)=-2.16\times 10^{-15}$ GeV and
$\Gamma_h(m_b)=3.61\times 10^{-14}$ GeV, which amounts to the branching fraction 
$B(b\to c\bar ud)=9.0\%$, given the total decay width of the 
$B^\pm$ meson $\Gamma^{\rm tot}=4.02\times 10^{-13}$ GeV \cite{PDG}. The dominant 
percent-level branching fractions of $B^+$ meson decays, 0.47\% from $\bar D^0\pi^+$,
1.34\% from $\bar D^0\rho^+$, 0.41\% from $\bar D^0\omega\pi^+$, 
0.56\% from $\bar D^0\pi^+\pi^+\pi^-$, 0.52\% from $\bar D^\star(2007)^0\pi^+$,
0.98\% from $\bar D^\star(2007)^0\rho^+$, 0.45\% from $\bar D^\star(2007)^0\omega\pi^+$,
1.03\% from $\bar D^\star(2007)^0\pi^+\pi^+\pi^-$,
1.8\% from $\bar D^\star(2007)^0\pi^-\pi^+\pi^+\pi^0$, 0.57\% from $\bar D^{\star 0}3\pi^+2\pi^-$ and
0.57\% from $\bar D^{\star\star 0}\pi^+$ \cite{PDG},
add up to 8.7\%, in agreement with our prediction.


A remark is in order. A bottom quark can also decay into light quarks through the $Q\to u\bar ud$
channel. It has been found that the dispersion relation for a fictitious heavy quark $Q$ decay 
into light final states determines only the charm quark mass. The larger bottom quark mass, even if 
it appears as one of the roots that vanishes the first derivative with respect to $\omega$, will not 
be selected due to the worse associated stability. In other words, the $Q\to u\bar ud$ mode is not 
efficient for the constraint of the bottom quark mass. We have elucidated in Sec.~II that an 
appropriate correlation function needs to be chosen for the purpose. 
For example, the $D^0$ meson decay width does not work for the determination of the charm quark mass
owing to the chirally suppressed dimension-six contribution.
This guideline also applies to analyses in lattice QCD and sum rules, which require suitable
correlation functions for extractions of physical observables.


\subsection{The $Q\to d\mu^+ \nu_\mu$ and $Q\to u\tau^- \bar\nu_\tau$ Decays}

The investigations on semileptonic and hadronic decay widths should return the same heavy quark mass 
from the viewpoint of the dynamical consistency, so the dispersion relations for the former can 
be utilized to constrain lepton masses. The weak annihilation contribution in semileptonic decays 
plays a role similar to the Pauli interference effect in hadronic decays for establishing a solution. 
Because the $e\nu_e$ channel with a tiny electron mass contains a negligible annihilation contribution, 
which does not provide a sufficient power correction, the $\mu \nu_\mu$ and 
$\tau \nu_\tau$ channels will be studied here. Equation~(\ref{68}) is proportional, in the 
limit $m_Q\to m_\ell$, to
\begin{eqnarray}
\Gamma^{(6)+(7)}_{s}(m_Q) \propto \frac{3}{2}\frac{(m_Q^2-m_\ell^2)^2m_\ell^2}{m_Q^3}
+6\bar\Lambda \frac{(m_Q^2-m_\ell^2)m_\ell^2}{m_Q^2},
\label{ms}
\end{eqnarray}
where the expansion $m_{H_Q}=m_Q+\bar\Lambda$ has been inserted and the net dimension-seven 
contribution has been approximated by $3-7m_\ell^2/m_Q^2\approx -4+O((m_Q^2-m_\ell^2)/m_Q^2)$ to 
give the second term. It is evident that the above expression follows the aforementioned 
helicity suppression; namely, the weak annihilation contribution diminishes with a lepton mass. 
A feature different from hadronic widths is that the dimension-six contribution is constructive 
in the present case, and the dimension-seven contribution is destructive. It turns out that
the $Q\to d\mu^+\nu_\mu$ decay width becomes negative in the low $m_Q$ region, though it is 
positive at $m_Q=m_c$, and some adjustment of the HQE input is necessary for solving the dispersion
relation. Therefore, we discuss the $Q \to u \tau^- \bar\nu_\tau$ decay first, whose width remains 
positive in the whole range of $m_Q$. 


As indicated in Eq.~(\ref{ms}), both the even piece $\Gamma_e^{\rm HQE}(m)$ and the odd piece 
$\Gamma_o^{\rm HQE}(m)$ have poles at $m= 0$. For the even piece, we consider the contour 
integration of the correlator $m^2\Pi(m)/(m^2-m_\tau^2)$ with the perturbative threshold 
$m_\tau$. The high-mass behavior stays the same, and 
the $m=0$ pole has been removed, so that Eq.~(\ref{con}) holds. The similar procedure
applied to Eq.~(\ref{mas}) yields 
\begin{eqnarray}
\int_{m_\tau^2}^\infty\frac{m^2\Delta\Gamma_e(m)}{(m^2-m_\tau^2)(m_Q^2-m^{2})}dm^2=0,\label{20}
\end{eqnarray}
where the unknown function $\Delta\Gamma_e(m)$ is fixed to $-\Gamma_e^{\rm HQE}(m)$ in the 
interval $(m_\tau,m_F)$ (a neutrino is regarded as being massless) with the physical threshold
$m_F=m_{\pi^0}+m_\tau$. For the odd piece, we consider the contour integration of the 
correlator $m^4\Pi(m)/(m^2-m_\tau^2)^2$ to remove the triple pole at $m=0$, and derive
the dispersion relation
\begin{eqnarray}
& &\int_{m_F}^R\frac{m^4\Gamma_e(m)}{(m^2-m_\tau^2)^2(m_Q-m)}dm-
\int^{-m_F}_{-R}\frac{m^4\Gamma_e(m)}{(m^2-m_\tau^2)^2(m_Q-m)}dm\nonumber\\
&=&\int_{m_\tau}^R\frac{m^4\Gamma_e^{\rm HQE}(m)}{(m^2-m_\tau^2)^2(m_Q-m)}dm-
\int^{-m_\tau}_{-R}\frac{m^4\Gamma_e^{\rm HQE}(m)}{(m^2-m_\tau^2)^2(m_Q-m)}dm.
\end{eqnarray}
We then have, with the variable change $m\to -m$ for the second integrals on both sides,
\begin{eqnarray}
\int_{m_\tau^2}^\infty\frac{m^3\Delta\Gamma_o(m)}{(m^2-m_\tau^2)^2(m_Q^2-m^{2})}dm^2=0,\label{21}
\end{eqnarray}
where the upper bound of the integration variable $m^2$ has been pushed to infinity. The unknown 
function $\Delta\Gamma_o(m)$ is fixed to $-\Gamma_o^{\rm HQE}(m)$ in the interval $(m_\tau,m_F)$.


The solutions from Eqs.~(\ref{20}) and (\ref{21}) are written as
\begin{eqnarray}
\frac{m^2\Delta \Gamma_e(m)}{m^2-m_\tau^2}&\approx&
y_e\left(\omega \sqrt{m^2-m_\tau^2}\right)^{\alpha} J_\alpha\left(2\omega\sqrt{m^2-m_\tau^2}\right),
\label{se2}\\
\frac{m^3\Delta \Gamma_o(m)}{(m^2-m_\tau^2)^2}&\approx&
y_o\left(\omega \sqrt{m^2-m_\tau^2}\right)^{\alpha} J_\alpha\left(2\omega\sqrt{m^2-m_\tau^2}\right),
\label{so2}
\end{eqnarray} 
respectively, whose combination gives the subtracted width of the $Q\to u\tau^-\bar\nu_\tau$ decay
\begin{eqnarray}
\Delta \Gamma_s(m_Q)\approx y_e
\left(1+\frac{y_o}{y_e} \frac{m_Q^2-m_\tau^2}{m_Q}\right)\frac{m_Q^2-m_\tau^2}{m_Q^2}
\left(\omega \sqrt{m_Q^2-m_\tau^2}\right)^{\alpha} 
J_\alpha\left(2\omega\sqrt{m_Q^2-m_\tau^2}\right).\label{so3}
\end{eqnarray}
Comparing the behaviors of Eqs.~(\ref{ms}) and (\ref{so3}) in the limit $m_Q\to m_\tau$,
we set the index $\alpha=0$ and the ratio 
\begin{eqnarray}
\frac{y_o}{y_e}=\frac{1}{4\bar\Lambda}=0.5\;{\rm GeV}^{-1}.
\end{eqnarray}
The boundary condition $\Delta \Gamma_s(m_F)=-\Gamma_s^{\rm HQE}(m_F)$ at $m_Q=m_F$ fixes the 
overall coefficient 
\begin{eqnarray}
y_e=-\Gamma_s^{\rm HQE}(m_F)\left[\left(1+\frac{m_F^2-m_\tau^2}{4\bar\Lambda m_F}\right)
\frac{(m_F^2-m_\tau^2)}{m_F^2}J_0\left(2\omega\sqrt{m_F^2-m_\tau^2}\right)\right]^{-1}.\label{dc4}
\end{eqnarray}


\begin{figure}
\begin{center}
\includegraphics[scale=0.35]{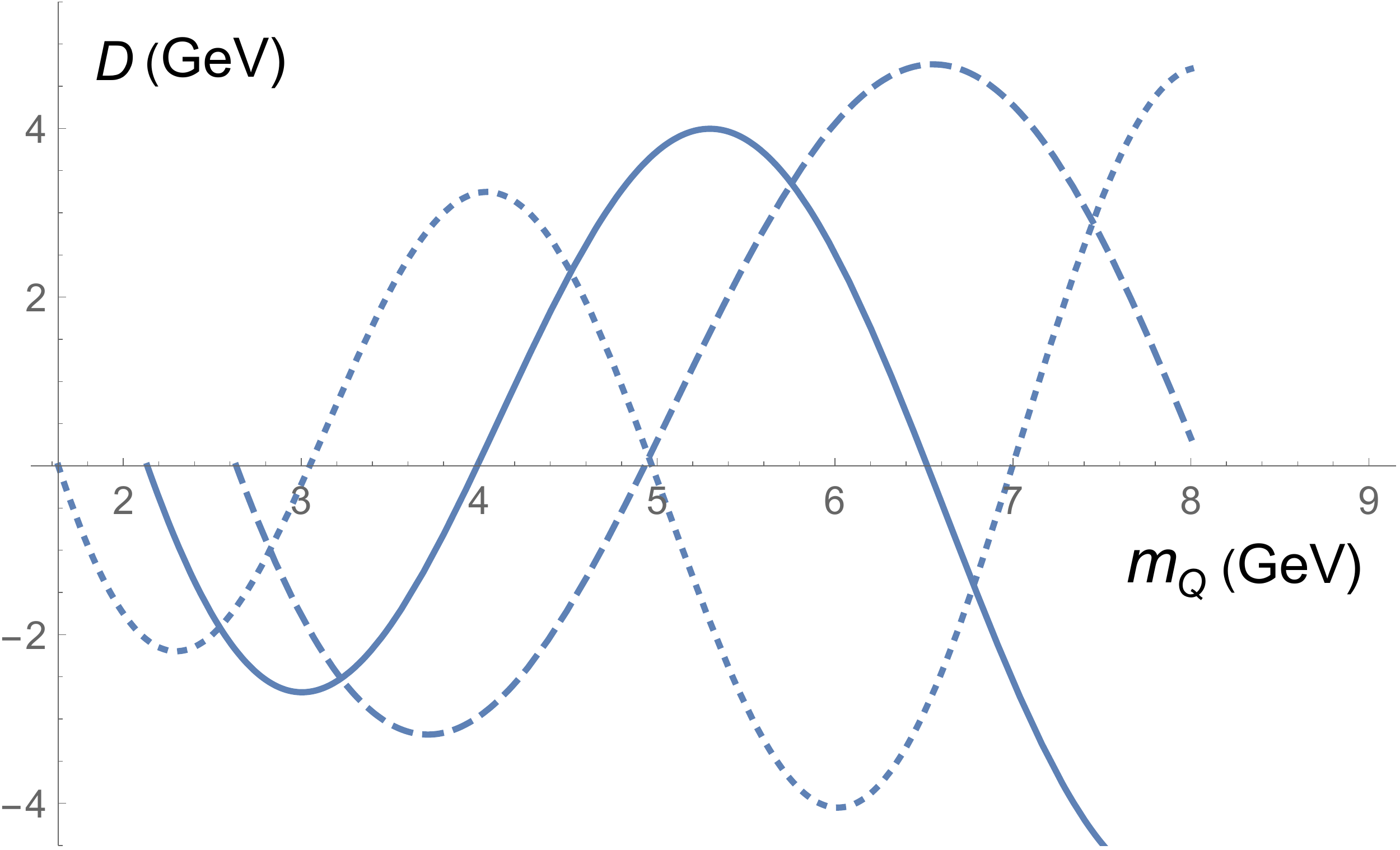}
\caption{\label{fig10} 
Dependencies of $D(m_Q)$ in Eq.~(\ref{der3}) on $m_Q$ for $m_\tau=1.5$ GeV
(dotted line), 2.0 GeV (solid line) and 2.5 GeV (dashed line).}
\end{center}
\end{figure}

The parameter $\bar\omega$ is obtained from the best fit of Eq.~(\ref{so3})
to $-\Gamma^{\rm HQE}_{s}(m_Q)$ in the interval $(m_\tau,m_F)$ with 
the same binding energy $\bar\Lambda=0.5$ GeV. We get $\bar\omega=0.748$ GeV$^{-1}$, 0.570 GeV$^{-1}$ 
and 0.460 GeV$^{-1}$ for the three different $\tau$ lepton masses $m_\tau=1.5$ GeV, 2.0 GeV and 2.5 GeV, 
respectively. We then search for the roots of the vanishing derivative
\begin{eqnarray}
D(m_Q)=\frac{d}{d\omega}\frac{J_0\left(2\omega\sqrt{m_Q^2-m_\tau^2}\right)}
{J_0\left(2\omega\sqrt{m_F^2-m_\tau^2}\right)}\Big|_{\omega=\bar\omega}=0,\label{der3}
\end{eqnarray}
where the $\omega$-independent factors in Eq.~(\ref{so3}) have been dropped.
The dependencies of the above derivative on $m_Q$ for the three values of $\bar\omega$ 
are exhibited in Fig.~\ref{fig10}, where the first roots located at $m_Q=m_F$ have no
physical significance, and the second roots read $m_Q=3.0$ GeV, 4.0 GeV and 4.9 GeV for 
$\bar\omega=0.748$ GeV$^{-1}$, 0.570 GeV$^{-1}$ and 0.460 GeV$^{-1}$ ($m_\tau=1.5$ GeV, 2.0 GeV 
and 2.5 GeV), respectively. It is observed that the $\tau$ lepton must have a mass around 2 GeV 
in order to produce the bottom quark mass $m_b= 4.03$ GeV, the same as that extracted from 
the $Q\to c u\bar d$ decay width. To be precise, the $\tau$ lepton takes the mass $m_\tau=2.02$ GeV. 
A lower (higher) $m_\tau$ would lead to a lower (higher) $m_b$. The value $m_\tau=2.02$ GeV, 
deviating from the measured one $m_\tau=1.777$ GeV \cite{PDG} by 14\%, is satisfactory enough in 
the current preliminary attempt. Following the prescription for estimating the uncertainty involved 
in the $Q\to su\bar d$ decay, we determine the $\tau$ lepton mass $m_\tau=(2.02\pm 0.02)$ 
GeV, where the error comes from the allowed range of $\bar\omega$. We remark that
the above analysis is independent of the effective gluon mass $m_g$ in the absence of the Wilson 
coefficients, so the constraint on $m_\tau$ is quite rigid. Similarly, we emphasize the nontrivial 
and stringent correlation among the concerned masses, instead of their exact values.

\begin{figure}
\begin{center}
\includegraphics[scale=0.3]{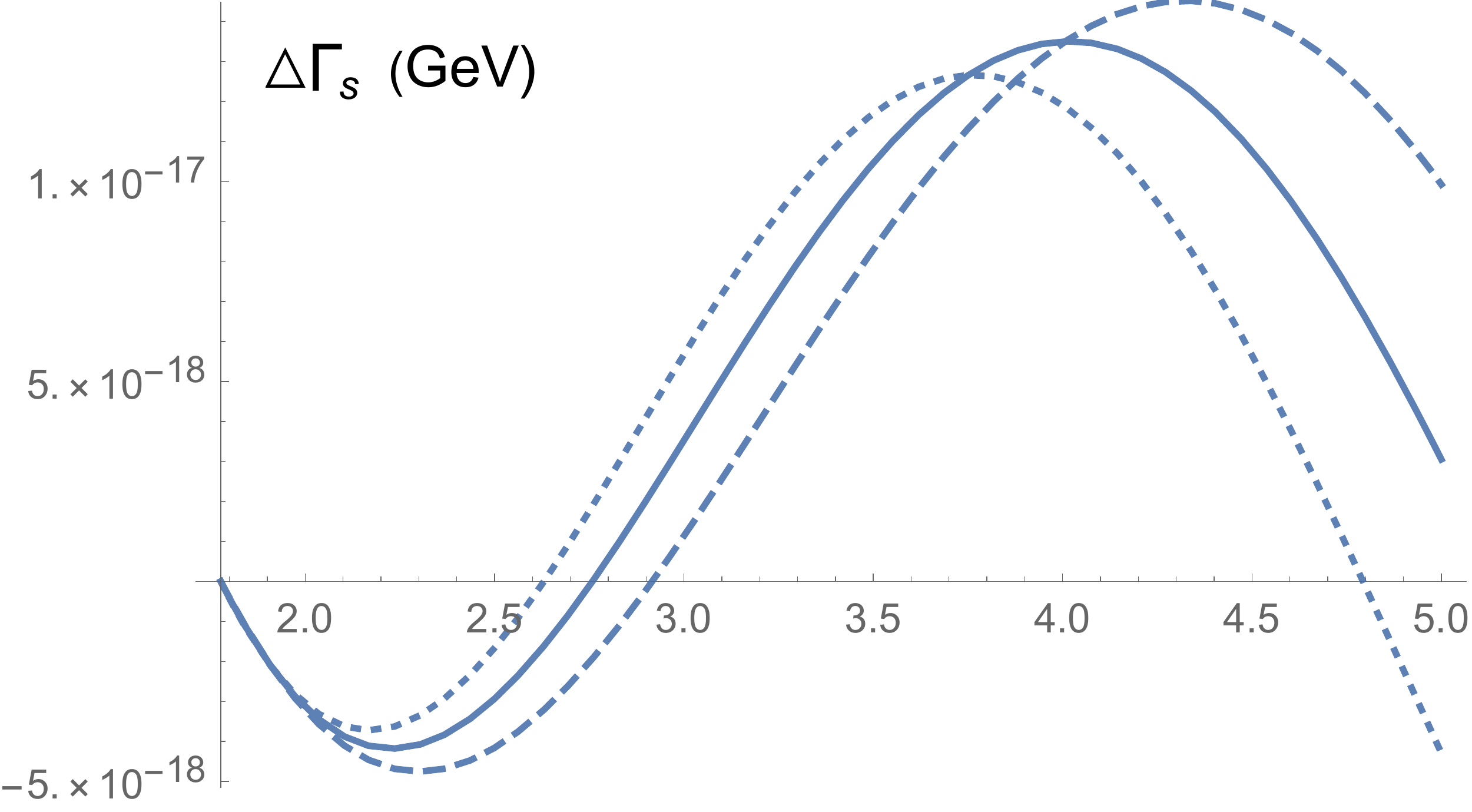}\hspace{1.0 cm} 
\includegraphics[scale=0.3]{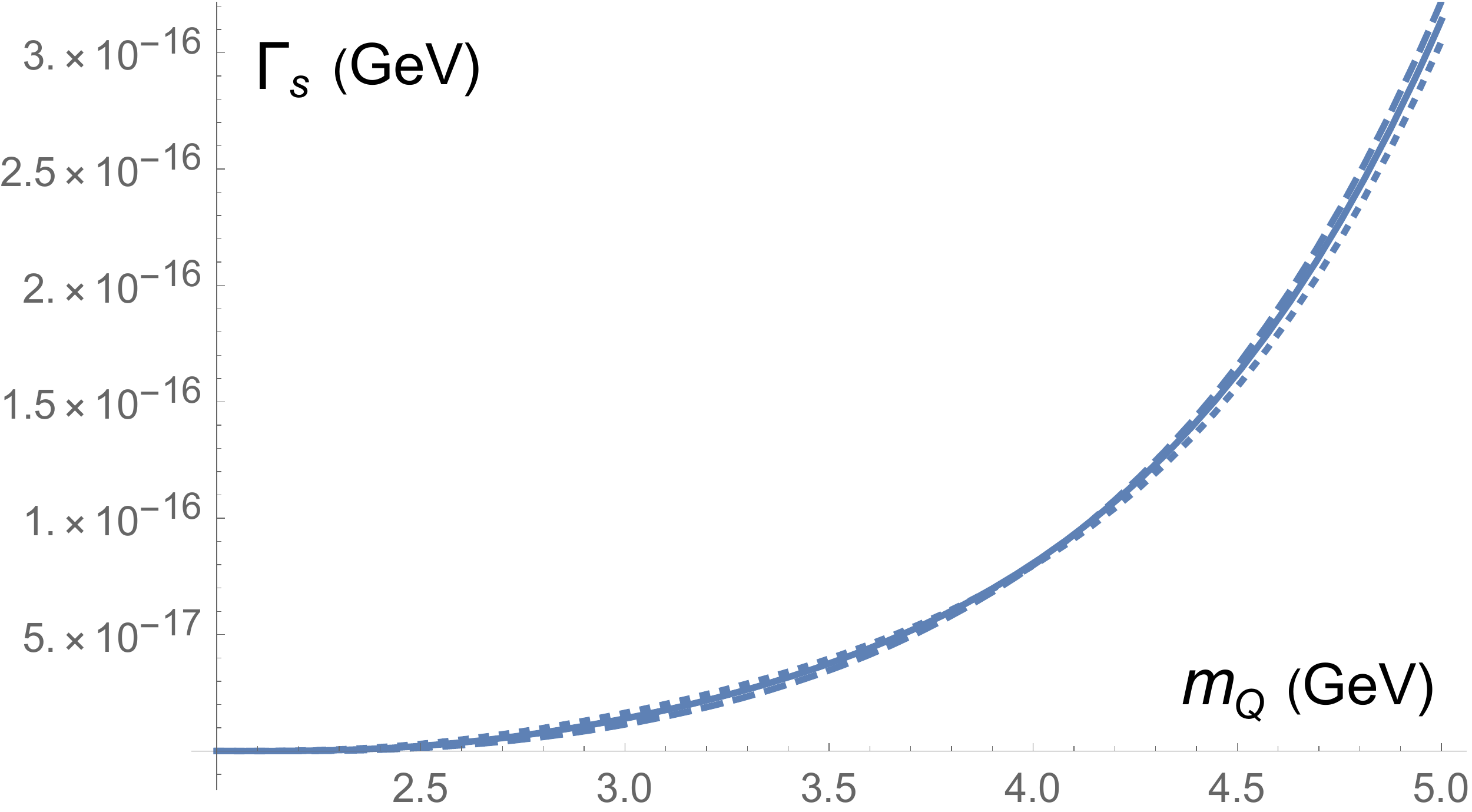}

(a) \hspace{7.0 cm} (b)
\caption{\label{fig11} 
Dependencies of (a) the subtracted width $\Delta\Gamma_s(m_Q)$ and (b) the width $\Gamma_s(m_Q)$ 
of the $Q\to u\tau^-\bar\nu_\tau$ decay on $m_Q$ for $\omega=0.52$ GeV$^{-1}$ (dashed line), 
0.570 GeV$^{-1}$ (solid line) and 0.62 GeV$^{-1}$ (dotted line).}
\end{center}
\end{figure}

The dependencies of the subtracted width $\Delta \Gamma_s(m_Q)$ and the
width $\Gamma_s(m_Q)$ on $m_Q$ are displayed for three values of $\omega$ around 
$\bar\omega=0.570$ GeV$^{-1}$, i.e., $\omega=0.52$ GeV$^{-1}$, 0.570 GeV$^{-1}$ and 0.62 GeV$^{-1}$,
in Figs.~\ref{fig11}(a) and \ref{fig11}(b), respectively. The CKM factor $|V_{\rm CKM}|^2=|V_{ub}|^2$ 
has been included with $|V_{ub}|=A\lambda^3\sqrt{\bar \rho^2 + \bar\eta^2}$, where the Wolfenstein 
parameters take the values $\bar\rho = 0.159$ and $\bar\eta = 0.348$ \cite{PDG}. The diminishing of 
$\Gamma_s(m_Q)$ for $m_Q> m_\tau$ up to $m_Q\approx 2.5$ GeV in 
Fig.~\ref{fig11}(b) certifies the superb coincidence between $\Delta\Gamma_s(m_Q)$ and 
$-\Gamma^{\rm HQE}_{s}(m_Q)$ in the interval $(m_\tau,m_F)$. All three curves 
cross each other in the small region located at $m_Q\approx 4$ GeV in Fig.~\ref{fig11}(a), suggesting 
the stability of the widths at this $m_Q$ under the variation of $\omega$. Though
the magnitude of $\Delta \Gamma_s(m_Q)$ is smaller than $\Gamma_s(m_Q)$, the intersection of 
the curves is still visible in Fig.~\ref{fig11}(b). We read off the decay widths 
at $m_b\approx 4.0$ GeV: $\Delta\Gamma_s(m_b)=1.65\times 10^{-17}$ GeV and
$\Gamma_s(m_b)=8.06\times 10^{-17}$ GeV, which also imply a minor nonperturbative contribution to 
$B$ meson decays. The above prediction for $\Gamma_s(m_b)$ is equivalent to the branching 
fraction $B(b\to u\tau^-\bar nu_\tau)=2.0\times 10^{-4}$, for which no data are available so far. 
However, it ought to be lower than the measured $b\to ue^-\bar\nu_e$ branching fraction 
$1.65\times 10^{-3}$ \cite{PDG}, so at least its order of magnitude makes sense.

\begin{figure}
\begin{center}
\includegraphics[scale=0.35]{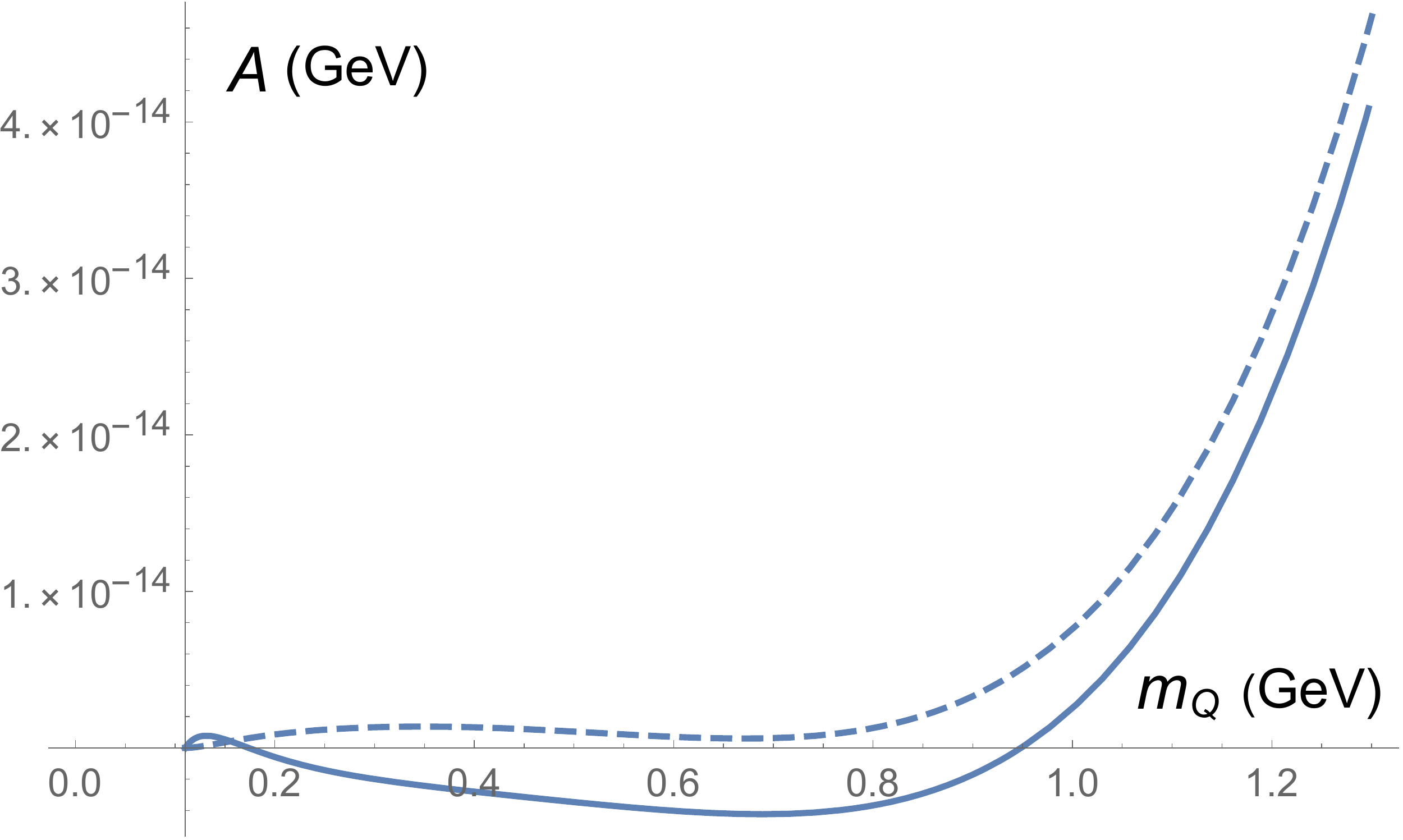}

\caption{\label{fig12}
The HQE widths of the semileptonic decay $Q\to d\mu^+\nu_\mu$ divided by the CKM factor,
$A(m_Q)\equiv \Gamma_s^{\rm HQE}(m_Q)/|V_{\rm CKM}|^2$, with (solid line) and without 
(dashed line) the dimension-seven contribution.}
\end{center}
\end{figure}

Next we constrain the muon mass $m_\mu$ from the dispersion relation for the semileptonic decay 
$Q\to d\mu^+\nu_\mu$. As stated before, the destructive dimension-seven contribution renders the HQE 
width slightly negative in the low $m_Q$ region. To be quantitative, we display the HQE width divided 
by the CKM factor, $A(m_Q)\equiv \Gamma_s^{\rm HQE}(m_Q)/|V_{\rm CKM}|^2$, with and without the 
dimension-seven contribution in Fig.~\ref{fig12}, where the muon mass has been set to $m_\mu=0.1$ GeV 
for illustration. The binding energy is kept as $\bar\Lambda=0.5$ GeV. It is seen that the former is 
negative in the range 0.15 GeV $< m_Q < 0.95$ GeV. Because the dimension-seven contribution is not 
yet complete, it is likely that this unphysical result would be amended in a full calculation. 
One possibility to go around this annoyance is to neglect the dimension-seven contribution, 
which is the least leading one in the HQE anyway. With the HQE input up to the dimension-six weak 
annihilation contribution, Eq.~(\ref{68}) becomes, in the limit $m_Q\to m_\ell$,
\begin{eqnarray}
\Gamma^{(6)+(7)}_{s}(m_Q) \propto \frac{3}{2}\frac{(m_Q^2-m_\ell^2)^2m_\ell^2}{m_Q^3}
\left(1+\frac{\bar\Lambda}{m_Q}\right),
\label{mm}
\end{eqnarray}
where the second term in the parentheses comes from the expansion $m_{H_Q}=m_Q+\bar\Lambda$
for the dimension-six piece. The similar procedure yields the solution
\begin{eqnarray}
\Delta \Gamma_s(m_Q)\approx y_e
\left(1+\frac{y_o}{y_e} m_Q\right)\frac{(m_Q^2-m_\mu^2)^2}{m_Q^4}
\left(\omega \sqrt{m_Q^2-m_\mu^2}\right)^{\alpha} 
J_\alpha\left(2\omega\sqrt{m_Q^2-m_\mu^2}\right),\label{db3}
\end{eqnarray}
which is fixed to $-\Gamma_s^{\rm HQE}(m_Q)$ in the interval $(m_\mu,m_F)$ of $m_Q$ with  
$m_F=m_{\pi^0}+m_\mu$.

Comparing Eqs.~(\ref{mm}) and (\ref{db3}) in the limit $m_Q\to m_\mu$,
we set the index $\alpha=0$ and the ratio 
\begin{eqnarray}
\frac{y_o}{y_e}=\frac{1}{\bar\Lambda}=2.0\;{\rm GeV}^{-1}.
\end{eqnarray}
The boundary condition at $m_Q=m_F$ specifies the overall coefficient
\begin{eqnarray}
y_e=-\Gamma_s^{\rm HQE}(m_F)
\left[\left(1+\frac{m_F}{\bar\Lambda}\right)\frac{(m_F^2-m_\mu^2)^2}{m_F^4}
J_0\left(2\omega\sqrt{m_F^2-m_\mu^2}\right)\right]^{-1}.\label{xta}
\end{eqnarray}
Matching Eq.~(\ref{db3}) to $-\Gamma_s^{\rm HQE}(m_Q)$ in the interval $(m_\mu,m_F)$ determines the 
parameter $\bar\omega=2.008$ GeV$^{-1}$, 1.475 GeV$^{-1}$ and 1.163 GeV$^{-1}$ for the three different
muon masses $m_\mu=0.08$ GeV, 0.11 GeV and 0.14 GeV, respectively. The fit does not involve the 
effective gluon mass $m_g$ in the absence of the Wilson coefficients.

\begin{figure}
\begin{center}
\includegraphics[scale=0.35]{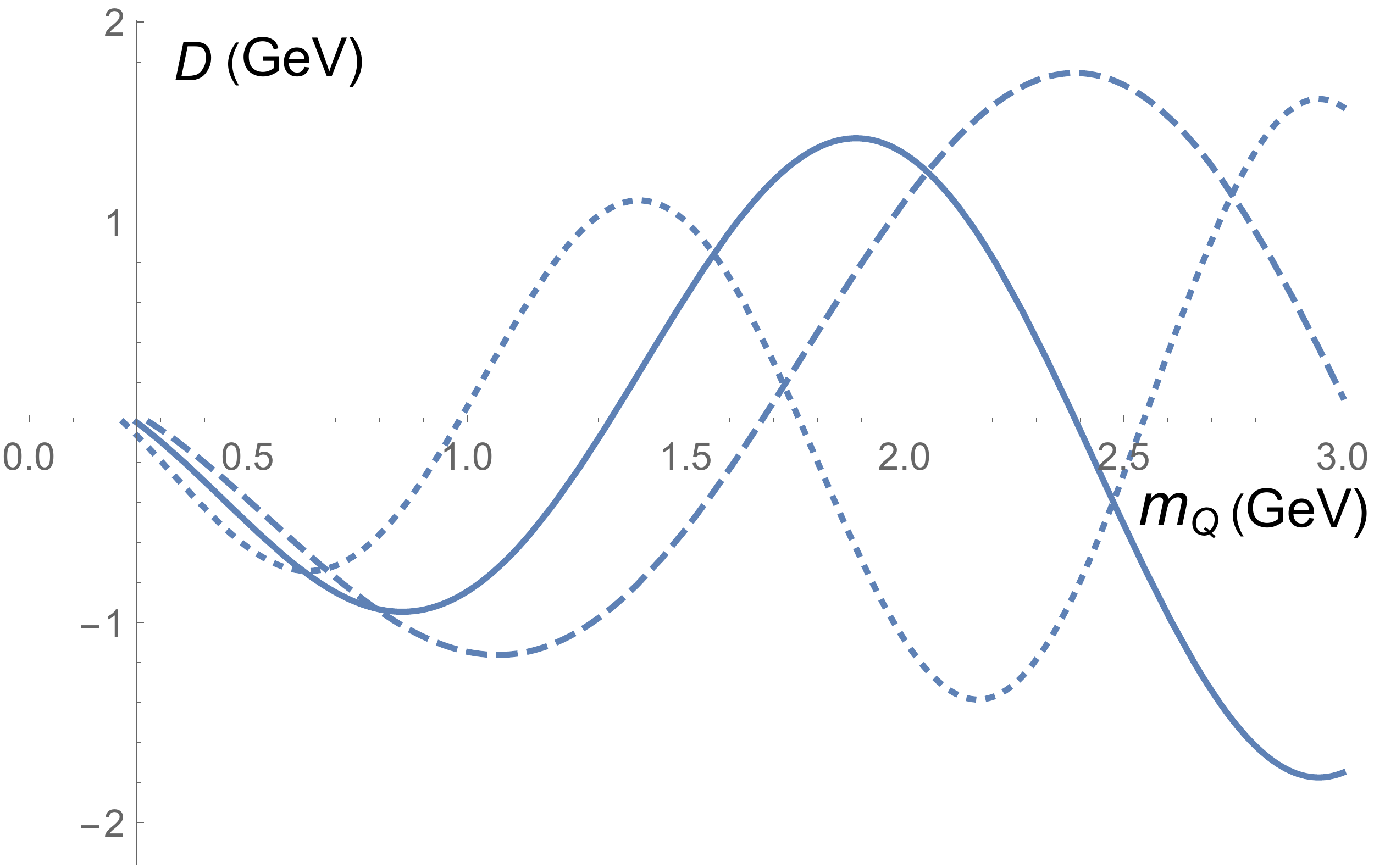}
\caption{\label{fig13} 
Dependencies of $D(m_Q)$ in Eq.~(\ref{der3}) on $m_Q$ for the $Q\to d\mu^+\nu_\mu$ decay
with $m_\mu=0.08$ GeV (dotted line), 0.11 GeV (solid line) and 0.14 GeV (dashed line).}
\end{center}
\end{figure}

The stability of the solved decay width requests the vanishing derivative in Eq.~(\ref{der3}) 
with $m_\tau$ being replaced by $m_\mu$. The dependence of the derivative on $m_Q$ is plotted in 
Fig.~\ref{fig13}, where the second roots $m_c=0.98$ GeV, $1.32$ GeV and 1.67 GeV corresponding to 
$m_\mu=0.08$ GeV, 0.11 GeV and 0.14 GeV, respectively, endow the decay width with the maximal 
stability under the variation of $\omega$. A lower (higher) lepton mass $m_\mu$ would
lead to a lower (higher) $m_c$. It is obvious that the muon mass $m_\mu=0.11$ GeV is preferred, 
because the resultant $m_c=1.32$ GeV is very close to 1.35 GeV extracted from the hadronic 
decay widths. We would like to make explicit that the choice
$m_\mu=0.112$ GeV reproduces $m_c=1.35$ GeV exactly. It is amazing that the determined muon mass 
deviates from the measured value 0.106 GeV \cite{PDG} by only about 6\%. It is also noticed that the 
locations of the roots are very sensitive to the values of $m_\mu$, supporting the effectiveness of 
the dispersive constraint on the lepton masses. Following the similar prescription for estimating 
the uncertainty involved in the $Q\to su\bar d$ decay, we determine
the muon mass $m_\mu=(0.112\pm 0.003)$ GeV, where the error comes from the allowed range of
$\bar\omega$.

\begin{figure}
\begin{center}
\includegraphics[scale=0.3]{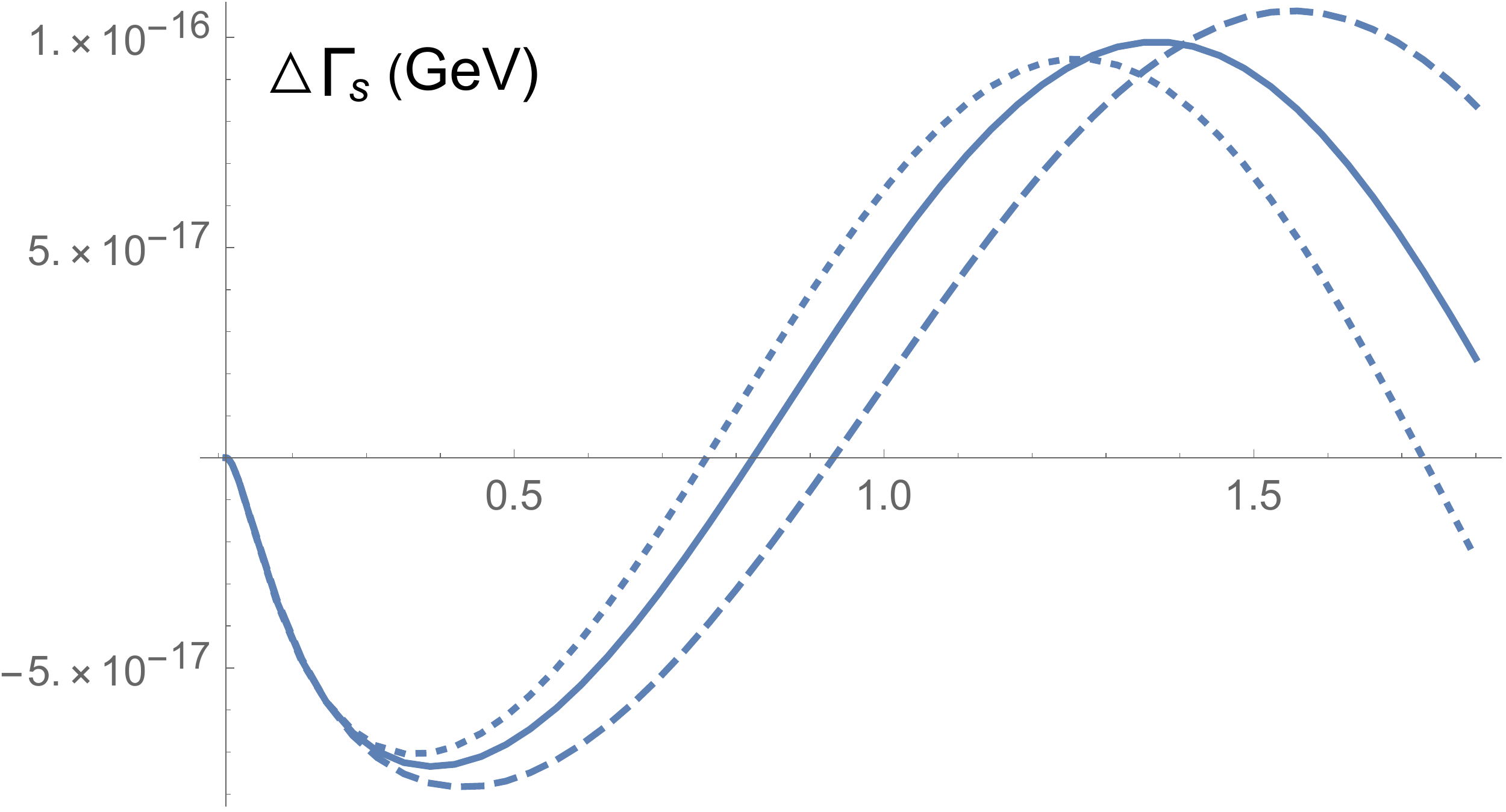}\hspace{1.0 cm} 
\includegraphics[scale=0.3]{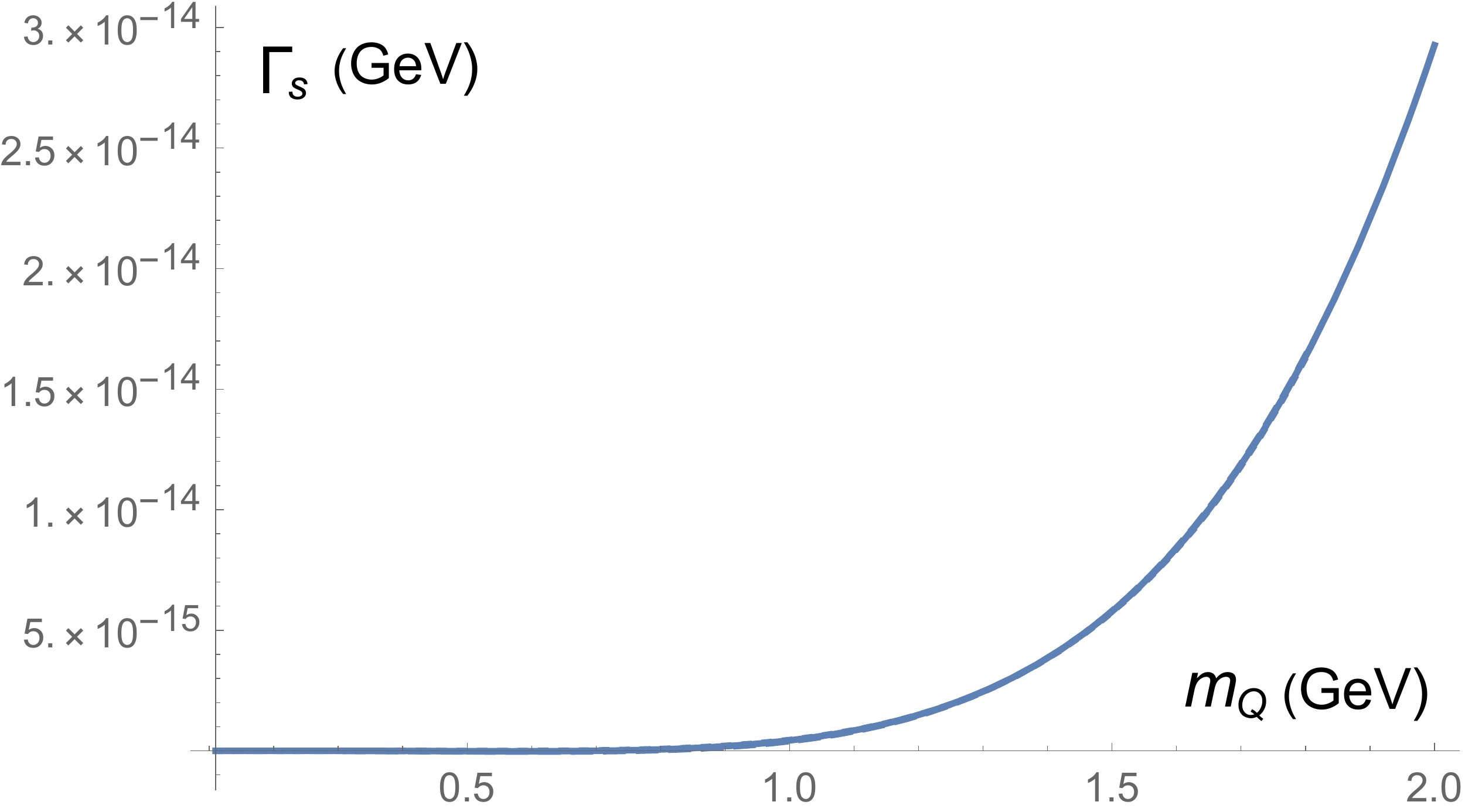}

(a) \hspace{7.0 cm} (b)
\caption{\label{fig14} 
Dependencies of (a) the subtracted width $\Delta\Gamma_s(m_Q)$ and (b) the width $\Gamma_s(m_Q)$ 
of the $Q\to d\mu^+\nu_\mu$ decay on $m_Q$ for $\omega=1.3$ GeV$^{-1}$ (dashed line), 
1.475 GeV$^{-1}$ (solid line) and 1.6 GeV$^{-1}$ (dotted line).}
\end{center}
\end{figure}

The dependencies of the subtracted width $\Delta \Gamma_s(m_Q)$ in Eq.~(\ref{db3}) and 
the width $\Gamma_s(m_Q)$ on $m_Q$ are exhibited in Figs.~\ref{fig14}(a) and \ref{fig14}(b),
respectively, for three values of $\omega$ around $\bar\omega$ from the best fit: $\omega=1.3$ 
GeV$^{-1}$, 1.475 GeV$^{-1}$ and 1.6 GeV$^{-1}$. The CKM factor $|V_{\rm CKM}|^2=|V_{cd}|^2$ has been 
included. The curves have a shape similar to that of the $Q\to u\tau^+\bar\nu_\tau$ decay. The 
diminishing $\Gamma_s(m_Q)$ for $m_Q> m_\mu$ up to $m_Q\approx 0.9$ GeV in 
Fig.~\ref{fig14}(b) evinces the perfect match between $\Delta\Gamma_s(m_Q)$ and 
$-\Gamma^{\rm HQE}_{s}(m_Q)$ in the interval $(m_\mu,m_F)$. The three curves 
cross each other more tightly at $m_Q\approx 1.3$-1.4 GeV in Fig.~\ref{fig14}(a), 
implying the stability of the widths evaluated around this $m_Q$ under the variation of $\omega$. 
Since the magnitude of $\Delta \Gamma_s(m_Q)$ is much lower than $\Gamma_s(m_Q)$, 
the three curves overlap almost exactly in Fig.~\ref{fig14}(b). We read off the decay widths 
at $m_c\approx 1.35$ GeV: $\Delta\Gamma_s(m_c)\approx 1\times 10^{-16}$ GeV and
$\Gamma_s(m_c)\approx 3\times 10^{-15}$ GeV. The negligible $\Delta \Gamma_s(m_Q)$ hints a tiny 
nonperturbative contribution to $D$ meson semileptonic decays, distinct from the case of hadronic 
decays. The predicted $\Gamma_s(m_c)$ amounts to the 
branching fraction $B(c\to d\mu^+\nu_\mu)\approx 5\times 10^{-3}$.
The sum of the branching fractions $3.50\times 10^{-3}$, $1.04\times 10^{-3}$, $2.4\times 10^{-3}$ 
and $1.77\times 10^{-3}$ for the $(\pi^0,\eta,\rho^0,\omega)\mu^+\nu_\mu$ modes, respectively, 
gives $8.71\times 10^{-3}$, not far from our prediction.

\section{CONCLUSION}

We have shown that dispersive analyses of physical observables can disclose stringent connections on 
high-energy and low-energy dynamics in the SM. It has been elaborated that the dispersion relations 
for inclusive heavy quark decay widths, whose definitions involve flavor-changing 
four-quark operators, correlate initial heavy quark masses and light final-state masses originating 
from the chiral symmetry breaking. The dispersion relation was reformulated in terms of the subtracted
decay width, i.e., the difference between the unknown width and the HQE width. A solution to the 
subtracted decay width was then constructed as an expansion of the generalized Laguerre polynomials, 
which satisfies the initial condition from the HQE input in the interval bounded by the quark- and 
hadron-level thresholds. Two arbitrary parameters were introduced into the formalism: the lowest 
degree $N$ for the polynomial expansion and the variable $\Lambda$, which scales the heavy quark mass 
in dispersive integrals. A solution to the dispersion relation should be insensitive to $\Lambda$, and 
this is possible only when a heavy quark takes a specific mass. A crucial feature is that the solution 
depends on the ratio $\omega^2=N/\Lambda$. Once the solution with a physical heavy quark mass is 
established, both $N$ and $\Lambda$ can be extended to arbitrarily large values by keeping $\omega$ in 
the stability window. All the large $N$ approximations assumed in solving the dispersion relation are 
then justified.

Starting with massless up and down quarks, we have determined the charm and bottom quark masses from 
the dispersion relations for the hadronic decays $Q\to du\bar d$ and $Q\to c\bar ud$, 
respectively. The strange quark, muon and $\tau$ lepton masses were constrained by the dispersion 
relations for the $Q\to s u\bar d$, $Q\to d\mu^+\nu_\mu$ and $Q\to u\tau^-\bar\nu_\tau$ decay widths, 
respectively, to generate the same parent heavy quark masses. It is clear that the chiral symmetry 
breaking plays an important role here, without which the initial interval bounded by the quark- and 
hadron-level thresholds shrinks, the solution for a subtracted decay width becomes identically zero, 
and no constraint can be imposed on the considered fermion masses. It has been scrutinized that the decay 
widths corresponding to the physical heavy quark masses agree with the available data. Except that the 
$Q\to du\bar d$ decay width is sensitive to the effective gluon mass, which is employed to stabilize 
the Wilson evolution to a low scale, the other channels are insensitive to the involved parameters. 
The observations made in this work are thus quite robust. We emphasize that no a priori information of 
a specific heavy quark was included: the fictitious mass $m_Q$ in the HQE expressions for the 
decay widths is arbitrary. All the inputs, such as the binding energy of a heavy meson, the 
HQET parameters and the effective gluon mass, take typical values. Therefore, the emergence of the 
charm and bottom quark masses in the stable solutions, their correlations with the strange quark, muon 
and $\tau$ lepton masses, and the consistency of the predicted decay widths with the data, are highly 
nontrivial.

Our goal is not to achieve an exact fit to the measured quantities, but to demonstrate 
that at least the fermion masses around the GeV scale, which characterize strong and weak 
dynamics, can be understood by means of the internal consistency of the SM, and that the dispersion 
relations for heavy quark decay widths strongly constrain those masses. No new symmetries 
or models beyond the SM, as attempted intensively in the literature, are needed. A more precise 
investigation can be invoked straightforwardly by taking into account subleading contributions to the 
HQE inputs, including those to the effective weak Hamiltonian, and more accurate HQET parameters and 
bag parameters. Though the present work is restricted to the fermion masses, the outcome has been 
convincing enough for conjecturing that other SM parameters can be also explained via the internal 
dynamical consistency, which wait for further exploration.


\section*{Acknowledgments}

We thank H.Y. Cheng for helpful discussions.
This work was supported in part by National Science and Technology Council of the Republic
of China under Grant No. MOST-110-2112-M-001-026-MY3.



\begin{thebibliography}{99}

\bibitem{FN79} C.~D.~Froggatt and H.~B.~Nielsen, 
Nucl. Phys. {\bf B147}, 277–298 (1979).


\bibitem{Feruglio:2017spp}
F.~Feruglio,
arXiv:1706.08749 [hep-ph].

\bibitem{Du:2022lij}
X.~K.~Du and F.~Wang,
JHEP \textbf{01}, 036 (2023).

\bibitem{Petcov:2022fjf}
S.~T.~Petcov and M.~Tanimoto,
arXiv:2212.13336 [hep-ph].

\bibitem{Kikuchi:2023cap}
S.~Kikuchi, T.~Kobayashi, K.~Nasu, S.~Takada and H.~Uchida,
Phys. Rev. D \textbf{107}, no.5, 055014 (2023).


\bibitem{Bree:2023ojl}
I.~Bree, S.~Carrolo, J.~C.~Romao and J.~P.~Silva,
Eur. Phys. J. C \textbf{83}, no.4, 292 (2023).

\bibitem{Abe:2023ilq}
Y.~Abe, T.~Higaki, J.~Kawamura and T.~Kobayashi,
arXiv:2301.07439 [hep-ph].

\bibitem{Seiberg:1994bz}
N.~Seiberg,
Phys. Rev. D \textbf{49}, 6857-6863 (1994).

\bibitem{Seiberg:1994pq}
N.~Seiberg,
Nucl. Phys. B \textbf{435}, 129-146 (1995).


\bibitem{Razamat:2020kyf}
S.~S.~Razamat and D.~Tong,
Phys. Rev. X \textbf{11}, 011063 (2021).

\bibitem{Hamada:2022ino}
Y.~Hamada and J.~Wang,
[arXiv:2209.15244 [hep-ph]].

\bibitem{Arkani-Hamed:1999ylh}
N.~Arkani-Hamed and M.~Schmaltz,
Phys. Rev. D \textbf{61}, 033005 (2000).

\bibitem{Giudice:2016yja}
G.~F.~Giudice and M.~McCullough,
JHEP \textbf{02}, 036 (2017).


\bibitem{Craig:2017cda}
N.~Craig, I.~Garcia Garcia and D.~Sutherland,
JHEP \textbf{10}, 018 (2017).


\bibitem{Li:2022jxc}
H.~n.~Li,
Phys. Rev. D \textbf{107}, no.5, 054023 (2023).

\bibitem{SVZ} M.~A.~Shifman, A.~I.~Vainshtein and V.~I.~Zakharov, Nucl. Phys. B {\bf 147}, 385 (1979);
B {\bf 147}, 448 (1979).

  
\bibitem{Li:2020xrz} 
  H.~n.~Li, H.~Umeeda, F.~Xu and F.~S.~Yu,
Phys. Lett. B \textbf{810}, 135802 (2020).

\bibitem{Xiong:2022uwj}
A.~S.~Xiong, T.~Wei and F.~S.~Yu,
arXiv:2211.13753 [hep-th].

\bibitem{Li:2021gsx}
H.~n.~Li,
Phys. Rev. D \textbf{104}, 114017 (2021).

\bibitem{Gratrex:2022xpm}
J.~Gratrex, B.~Meli\'c and I.~Ni\v{s}and\v{z}i\'c,
JHEP \textbf{07}, 058 (2022).


\bibitem{Blum:1999xi}
T.~Blum, A.~Soni and M.~Wingate,
Phys. Rev. D \textbf{60}, 114507 (1999).

\bibitem{Dominguez:2007my}
C.~A.~Dominguez, N.~F.~Nasrallah, R.~Rontsch and K.~Schilcher,
JHEP \textbf{05}, 020 (2008).

\bibitem{HFLAV:2022pwe}
Y.~Amhis \textit{et al.} [HFLAV],
Phys. Rev. D \textbf{107}, no.5, 052008 (2023).

\bibitem{Cheng:2018rkz}
H.~Y.~Cheng,
JHEP \textbf{11}, 014 (2018).

\bibitem{Li:2020fiz}
H.~n.~Li and H.~Umeeda,
Phys. Rev. D \textbf{102}, no.9, 094003 (2020).

\bibitem{Li:2020ejs}
H.~n.~Li and H.~Umeeda,
Phys. Rev. D \textbf{102}, 114014 (2020).

\bibitem{Li:2022qul}
H.~n.~Li,
Phys. Rev. D \textbf{106}, no.3, 034015 (2022).

\bibitem{Lenz:2013aua}
A.~Lenz and T.~Rauh,
Phys. Rev. D \textbf{88}, 034004 (2013).

\bibitem{KSUV} V.~A.~Khoze, M.~A.~Shifman, N.~G.~Uraltsev and M.~B.~Voloshin, Sov. J. Nucl. 
Phys. {\bf 46}, 112 (1987) [Yad. Fiz. {\bf 46}, 181 (1987)].

\bibitem{CGG} J.~Chay, H.~Georgi and B.~Grinstein, Phys. Lett. B {\bf 247}, 399 (1990).

\bibitem{Lenz:2022rbq}
A.~Lenz, M.~L.~Piscopo and A.~V.~Rusov,
[arXiv:2208.02643 [hep-ph]].

\bibitem{Bernlochner:2022ucr}
F.~Bernlochner, M.~Fael, K.~Olschewsky, E.~Persson, R.~van Tonder, K.~K.~Vos and M.~Welsch,
JHEP \textbf{10}, 068 (2022).


\bibitem{Bigi:1992su}
I.~I.~Y.~Bigi, N.~G.~Uraltsev and A.~I.~Vainshtein,
Phys. Lett. B \textbf{293}, 430-436 (1992)
[erratum: Phys. Lett. B \textbf{297}, 477-477 (1992)].

\bibitem{Falk:1994gw}
A.~F.~Falk, Z.~Ligeti, M.~Neubert and Y.~Nir,
Phys. Lett. B \textbf{326}, 145-153 (1994).

\bibitem{Mannel:2017jfk}
T.~Mannel, A.~V.~Rusov and F.~Shahriaran,
Nucl. Phys. B \textbf{921}, 211-224 (2017).

\bibitem{QP83} H.~K.~Quang and X.~Y.~Pham, Phys. Lett. B {\bf 122}, 297 (1983).
\bibitem{N89} Y.~Nir, Phys. Lett. B {\bf 221}, 184 (1989).

\bibitem{Bagan:1994zd}
E.~Bagan, P.~Ball, V.~M.~Braun and P.~Gosdzinsky,
Nucl. Phys. B \textbf{432}, 3-38 (1994).

\bibitem{Bagan:1994qw}
E.~Bagan, P.~Ball, V.~M.~Braun and P.~Gosdzinsky,
Phys. Lett. B \textbf{342}, 362-368 (1995)
[erratum: Phys. Lett. B \textbf{374}, 363-364 (1996)].

\bibitem{Bagan:1995yf}
E.~Bagan, P.~Ball, B.~Fiol and P.~Gosdzinsky,
Phys. Lett. B \textbf{351}, 546-554 (1995).

\bibitem{Neubert:1996we}
M.~Neubert and C.~T.~Sachrajda,
Nucl. Phys. B \textbf{483}, 339-370 (1997).

\bibitem{Uraltsev:1996ta}
N.~G.~Uraltsev,
Phys. Lett. B \textbf{376}, 303-308 (1996).

\bibitem{Beneke:1996xe}
M.~Beneke and G.~Buchalla,
Phys. Rev. D \textbf{53}, 4991-5000 (1996).

\bibitem{Franco:2002fc}
E.~Franco, V.~Lubicz, F.~Mescia and C.~Tarantino,
Nucl. Phys. B \textbf{633}, 212-236 (2002).

\bibitem{Beneke:2002rj}
M.~Beneke, G.~Buchalla, C.~Greub, A.~Lenz and U.~Nierste,
Nucl. Phys. B \textbf{639}, 389-407 (2002).


\bibitem{King:2021xqp}
D.~King, A.~Lenz, M.~L.~Piscopo, T.~Rauh, A.~V.~Rusov and C.~Vlahos,
JHEP \textbf{08}, 241 (2022).

\bibitem{Falk:2004wg}
  A.~F.~Falk, Y.~Grossman, Z.~Ligeti, Y.~Nir and A.~A.~Petrov,
  Phys.\ Rev.\ D {\bf 69}, 114021 (2004).


\bibitem{BBC} D.~Borwein, J.~M.~Borwein, R.~E.~Crandall, 
SIAM J. Numer. Anal. {\bf 46}, 3285–3312 (2008).

\bibitem{Kirk:2017juj}
M.~Kirk, A.~Lenz and T.~Rauh,
JHEP \textbf{12}, 068 (2017)
[erratum: JHEP \textbf{06}, 162 (2020)].

\bibitem{King:2021jsq}
D.~King, A.~Lenz and T.~Rauh,
JHEP \textbf{06}, 134 (2022).

\bibitem{Buchalla:1995vs}
G.~Buchalla, A.~J.~Buras and M.~E.~Lautenbacher,
Rev. Mod. Phys. \textbf{68}, 1125-1144 (1996).

\bibitem{Zhong:2021epq}
T.~Zhong, Z.~H.~Zhu, H.~B.~Fu, X.~G.~Wu and T.~Huang,
Phys. Rev. D \textbf{104}, 016021 (2021).

\bibitem{Aguilar:2015bud}
A.~C.~Aguilar, D.~Binosi and J.~Papavassiliou,
Front. Phys. (Beijing) \textbf{11}, no.2, 111203 (2016).

\bibitem{Gomez:2016xjz}
J.~D.~Gomez and A.~A.~Natale,
Int. J. Mod. Phys. A \textbf{32}, no.02n03, 1750012 (2017).

\bibitem{PDG}
R.L. Workman et al. (Particle Data Group), Prog. Theor. Exp. Phys. 2022, 083C01 (2022).




\end{thebibliography}
\end{document}